\pdfoutput=1
\documentclass{iopjournal}

\usepackage[utf8]{inputenc}
\usepackage[T1]{fontenc}
\usepackage{lmodern}
\usepackage{amsmath,amssymb}
\usepackage{float}
\usepackage{placeins}
\usepackage{multirow}
\usepackage[disable]{todonotes}

\renewcommand{\articletype}[1]{{\vspace*{-8mm}\noindent \large \sf Preprint submitted to Superconductor Science and Technology}\par\vspace*{8mm}}
\begin{document}

\articletype{Paper}

\title{Tailored V$_x$O$_y$ Metal--Insulator-Transition Oxides for Temperature-Triggered Smart Insulation in Superconducting Coils}

\author{M.~Ortino$^{1}$, D.~Uglietti$^{2}$, A.~Testino$^{3}$, N.~Casati$^{3}$, A.~Steppke$^{3}$, W.~Wnetkowska$^{3}$, A.~Masi$^{4}$ and K.~Sedlak$^{2}$}

\affil{$^{1}$GREEN, Universit\'e de Lorraine, Nancy, France}
\affil{$^{2}$Swiss Plasma Center, EPFL, Lausanne, Switzerland}
\affil{$^{3}$Paul Scherrer Institut (PSI), Villigen, Switzerland}
\affil{$^{4}$ENEA, Frascati, Italy}

\email{mattia.ortino@univ-lorraine.fr}

\keywords{smart insulation, metal-insulator transition, vanadium oxide, non-insulated coil, partially-insulated coil, quench protection, HTS, superconducting magnet}

\begin{abstract}
Smart insulation is a turn-to-turn layer that insulates in normal
operation and conducts only near a hot spot. It could combine the fast
ramping of insulated coils with the passive energy dump of
non-insulated coils. Metal--insulator transition (MIT) oxides are
natural candidates, but the only one tested in a coil so far, V$_2$O$_3$,
switches at 150--170~K, far above the 30--70~K window useful for REBCO
magnets. We explore the vanadium--oxygen system to lower the
transition. TGA redox processing produced multi-phase powders with an
effective transition near 127~K, about 25~K below V$_2$O$_3$.
Hydrothermal V$_3$O$_7$, hydrated or annealed, was far too resistive
and showed no usable transition. V$_2$O$_3$--epoxy specimens provided a
robust reference and an acceptance criterion. Cryogenic measurements
revealed a reproducible thermal hysteresis, and the measured specific
heat of V$_2$O$_3$ corrects earlier model inputs by close to two orders
of magnitude. A first electro-thermal model reproduces current sharing
through the smart-insulation layer. Doped V$_2$O$_3$ and V$_8$O$_{15}$ are
identified as the next targets.
\end{abstract}


\section{Introduction}
\label{sec:intro}

\subsection{Protecting magnets with high stored energy}
\label{sec:intro-nipi}

When a superconducting magnet quenches, part of its winding becomes
resistive. The magnetic energy stored in the coil, $E_{\mathrm{st}}$, is
then converted into heat. A first estimate of the final temperature
$T_{\max}$ assumes that this energy is spread uniformly over the
conductor mass $m$ \cite{wilson1983}:
\begin{equation}
  \int_{T_0}^{T_{\max}} c_p(T)\,\mathrm{d}T \;=\; \frac{E_{\mathrm{st}}}{m},
  \label{eq:tmax-estimate}
\end{equation}
where $c_p$ is the specific heat and $T_0 = 4.2$~K. As a rule of thumb,
about 2~J/g heats a winding to 50~K, 10~J/g to 100~K and 60~J/g to
300~K \cite{wilson1983}. A small laboratory solenoid (250~kJ, 50~kg)
stores about 5~J/g. A dipole of the Large Hadron Collider (LHC) and the
Compact Muon Solenoid (CMS) detector magnet store about 10--11~J/g.
Toroidal-field (TF) coils for fusion -- ITER, the EU DEMO baseline or
the SPARC TF model coil -- reach 20 to more than 50~J/g. At these
energy densities the winding cannot simply absorb its own stored energy
without damage.

For coils in the multi-GJ, tens-of-J/g class, two protection options
exist. In the first, the coil is insulated: the quench is detected and
the stored energy is dumped into an external resistor. In the second,
the coil is non-insulated (NI), for example through discrete
turn-to-turn bridges, and the energy is dumped passively inside the
winding. The two options differ clearly in ramp rate. An insulated coil
can be charged at any rate. An NI coil can only be ramped very slowly,
because current leaks through the turn-to-turn contacts. The SPARC
tokamak uses both. Its central solenoid, ramped at a few T/s, is
insulated \cite{hartwig2020}. Its TF coils, ramped very slowly
($\ll$1~mT/s), are non-insulated \cite{hartwig2024_tfmc}. In particle
accelerators, large detector magnets can use the second option, while
dipoles must use the first.

Discrete metal bridges between turns and between layers were studied as
an NI-like option for the EU DEMO TF winding pack
\cite{ortino2024_tasc,ortino2025_tasc_mit}. Their resistance is fixed
by construction, so they act everywhere, not only where a hot spot
forms. In an unpublished study (A.~Rossi, SPC/EPFL, 2022), welded
stainless-steel turn-to-turn bridges reached their target resistance
reliably. Layer-to-layer bridges, which need a much larger contact
area, did not. Graphite-filled cryogenic epoxies were two to three
orders of magnitude too resistive and drifted over repeated 77~K
cycles. Silver-filled epoxies were too conductive and cracked on
cooling. No fixed-resistivity material covered the required range.

\begin{figure}[H]
	\centering
	\includegraphics[width=\textwidth]{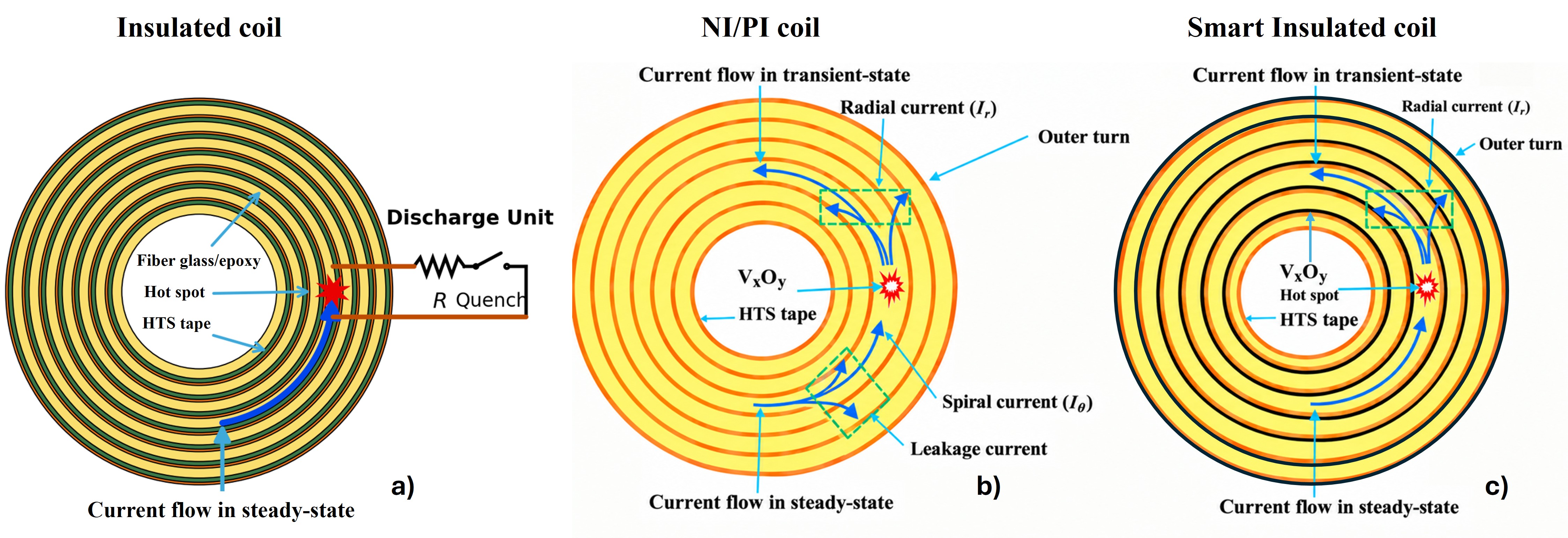}
	\caption{Coil-protection concepts for a single pancake winding:
		(a) conventional fibreglass/epoxy and/or polyimide insulation,
		relying on an external discharge unit; (b) non-insulated (NI),
		with always-on radial and spiral leakage paths; (c) smart
		insulation, insulating in normal operation and locally conductive
		only at the hot spot. Design partially inspired by
		\cite{mussa2024doped}.}
	\label{fig:pancake-concepts}
\end{figure}

An insulated coil whose turn-to-turn current paths switch on only
locally -- exactly where and when the temperature rises -- would be a
third option. It would combine the fast ramping of insulated coils with
the passive internal energy dump of NI coils. A smart-insulation (SI)
material could realise this third option. Figure~\ref{fig:pancake-concepts}
compares the three concepts for a single pancake winding. In a
conventionally insulated coil (a), the insulation blocks all
turn-to-turn current, and a hot spot can only be relieved by an
external discharge. In an NI coil (b), current can bypass a hot spot
through radial and spiral leakage paths, but these paths are also
active in normal operation. An SI layer (c) behaves like (a) in normal
operation and like (b) only near a hot spot.

\subsection{Smart-insulation materials}
\label{sec:intro-strategies}

An SI material is an insulator whose resistivity responds strongly to a
local quantity -- temperature, electric field or pressure. It insulates
in normal operation and conducts once a local threshold is exceeded.
Three material families have been considered.

\textbf{Percolative carbon-based composites} disperse a conductive
filler (carbon black, carbon nanotubes, graphene or short carbon
fibres) in an insulating polymer. Near the percolation threshold, the
ratio between room-temperature and liquid-helium conductivity can reach
$10^2$--$10^4$ \cite{bhatia2018_review}. In that regime, however,
conductivity is also most sensitive to filler content, dispersion and
processing, which a large winding cannot tolerate. Safely above
threshold, the switching ratio collapses to less than a factor of three
in well-optimised nanotube composites \cite{bhatia2018_weakT}, and to
one or two orders of magnitude in less optimised ones
\cite{bhatia2018_review}. The resistance is set by a geometric network
at fabrication, rather than switching at a defined temperature. In
addition, the polymer matrix is the radiation-sensitive component:
radiation-hardening work for magnet insulation embeds inorganic oxide
nanoparticles in such polymers for this reason \cite{guvvala2026_monap}.
A fully inorganic material is likely to tolerate fusion-relevant
neutron and gamma fluences better.

\textbf{Varistors} (ZnO- or SiC-based) are non-linear resistors
controlled by the local electric field. They are established in
external quench protection, where they replace linear dump resistors to
reduce peak voltage or hot-spot temperature
\cite{galvin_varistor_2022,galvin2022_hotspot,galvin2023_sic4k}.
Varistor-like turn-to-turn insulation has also been proposed: radial
current bypass starts once the local voltage across the layer exceeds a
threshold \cite{galvin_varistor_2022}. For the same hot-spot
temperature, the voltage in an HTS (high-temperature superconductor)
coil is lower than in an LTS (low-temperature superconductor) coil,
because at 30--40~K REBCO can still carry some current. Depending on
the threshold of the selected material, both triggers -- electric field
or temperature -- could work. For REBCO, the balance more likely tips
toward a temperature trigger.

\textbf{Metal--insulator transition (MIT) materials} switch between an
insulating low-temperature state and a metallic high-temperature state
at a transition temperature $T_{\mathrm{MIT}}$. The resistivity change
typically spans four to six orders of magnitude. The trigger is the
local temperature itself. A hot spot heats the adjacent MIT layer above
$T_{\mathrm{MIT}}$, current bypasses the hot region, and the layer
becomes insulating again once the winding cools down. The mechanism is
local, passive and reversible, and needs no detection or switching
hardware. This is why MIT materials are the focus of this work.

To be effective in a REBCO magnet, the transition should lie between
30~K and 70~K. Large REBCO magnets may operate at up to 20~K, with
transients up to 25--30~K. Above 70--80~K, the quench may have
progressed too far to keep the hot spot below 200~K.

The MIT switch is not instantaneous. The layer must first be heated
across a finite temperature interval, which is a conduction-limited
process. Interface-scale modelling in this study places this response
at about 0.5--2~s for representative hot-spot energies. HTS quenches
evolve on a comparable timescale. The normal-zone propagation velocity
(NZPV) of REBCO tape is typically only 1--100~cm/s \cite{ortino2024_tasc}.
The normal zone stays small and local for a long time. This is also why
voltage-based quench detection is difficult in HTS magnets, and why the
field is moving toward temperature-based detection, e.g.\ with
fibre-optic sensors \cite{salazar2021_fiberoptic}. A
temperature-triggered SI layer fits naturally with this trend.

LTS magnets are a different case. For Nb--Ti and Nb$_3$Sn the useful
window would be lower, about 8--30~K. Their quench dynamics are much
faster: NZPV is 10--80~m/s in Nb--Ti and 1--50~m/s in Nb$_3$Sn
\cite{ortino2024_tasc}. Simulating the interplay between quench and SI
layer is therefore more demanding, as it needs a much finer time and
space resolution. More importantly, a normal zone growing by metres per
second outruns a layer that needs about a second to switch. Full
self-protection of an LTS magnet by an SI layer therefore appears
excluded. Voltage-based quench detection, which works well for LTS,
remains necessary, and an SI layer could at most complement it. This
paper therefore focuses on HTS.

\subsection{V$_2$O$_3$: proof of concept and its limit}
\label{sec:intro-v2o3}

About seventy to one hundred MIT compounds are known
\cite{brahlek2017,hu2023}. V$_2$O$_3$ is the only one demonstrated so far
inside a superconducting coil. A series of 2G-HTS ``smart insulation''
coils used V$_2$O$_3$ powder mixed with a polymer binder and applied as
a paste between turns
\cite{kim2017_mit,jo2017_smartinsulation,kim2019_racetrack,kim2022_racetrack300}.
These coils showed resistivity changes of $10^5$--$10^6$ across the
transition. The transition appeared at about 150~K on cooling and 170~K
on heating; this 20~K hysteresis reflects the first-order character of
the transition. Other deposition routes have been explored with the
same compound: reactive-arc-deposited VO$_x$ coatings on REBCO tape
\cite{yang2020_voxcoating} and dip-coated V$_2$O$_3$ interlayers between
stacked tapes \cite{bonura2023contact}. The concept has also been
proposed for thin-walled detector magnets \cite{vaskuri2023}.

\begin{figure}[H]
  \centering
  \includegraphics[width=0.65\textwidth]{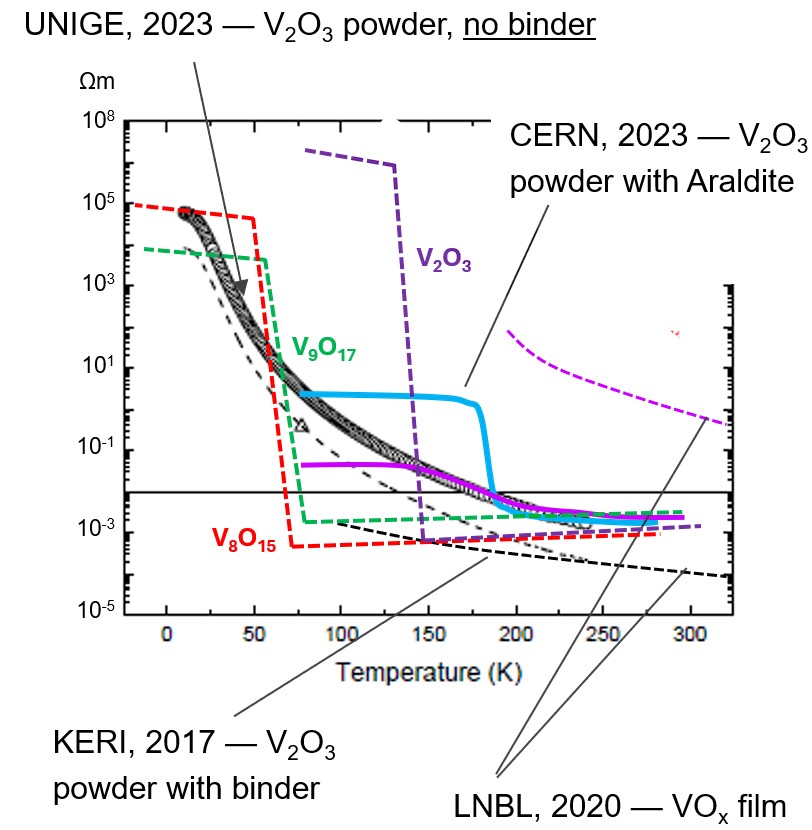}
  \caption{Resistivity vs.\ temperature for V$_2$O$_3$ from several
    groups and processing routes (UNIGE: powder, no binder
    \cite{bonura2023contact}; CERN: powder with Araldite \cite{vaskuri2023};
    KERI: powder with binder \cite{kim2017_mit}; LNBL: VO$_x$ film),
    compared with a V$_8$O$_{15}$ literature curve \cite{kawazoe2023_v8o15}.}
  \label{fig:lit-v2o3}
\end{figure}

The limit is equally clear. V$_2$O$_3$ switches at 150--170~K
\cite{kim2019_racetrack}, far above the 30--70~K window. Figure~\ref{fig:lit-v2o3}
collects V$_2$O$_3$ resistivity curves from several processing routes,
together with a V$_8$O$_{15}$ curve \cite{kawazoe2023_v8o15}. All
V$_2$O$_3$ curves switch at 150--170~K, whatever the processing route.
V$_8$O$_{15}$ switches at a much lower temperature. Moving the transition
into the useful window is the central materials problem of this paper.

\subsection{Why the vanadium--oxygen system}
\label{sec:intro-vxoy}

Other MIT compounds already transition closer to the target window
(Table~\ref{tab:mit-landscape}). Two of them, Al-doped ZnO and PYCCO,
(Pr$_{1-y}$Y$_y$)$_{1-x}$Ca$_x$CoO$_3$, are even tunable by composition
\cite{fisher2016_systematics,mit_review_2013_TODO}. Several drawbacks,
however, make them less attractive for a coil layer. Non-oxide
compounds such as SmRu$_4$P$_{12}$, FeSi or BaVS$_3$
\cite{gardner1969,matsuura1991,fazekas2007,tanaka2007} need rather
complex synthesis routes. Some phases rely on scarce or expensive
elements, such as iridium or rare earths, whereas vanadium is widely
distributed in the Earth's crust. Fe$_3$O$_4$ is ferrimagnetic, which
adds forces and stray fields inside a high-field winding. Finally, each
of these compounds is unrelated to the next: adopting one means
starting synthesis, processing and cryogenic testing from scratch.

\begin{table}[htbp]
  \centering
  \caption{Selected MIT compounds and reported transition temperatures.
    Upper block: diversity of MIT chemistries. Lower block: the V--O
    system alone. Sources:
    \cite{fisher2016_systematics,mit_review_2013_TODO,brahlek2017,hu2023}.}
  \label{tab:mit-landscape}
  \begin{tabular}{lcp{7.2cm}}
    \hline
    Compound & $T_{\mathrm{MIT}}$ (K) & Notes \\
    \hline
    SmRu$_4$P$_{12}$        & $\approx$16       & filled skutterudite; rare earth \\
    FeSi                     & $\approx$20       & non-oxide \\
    Nd$_2$Ir$_2$O$_7$        & $\approx$33       & pyrochlore iridate; rare earth and Ir \\
    Al:ZnO                   & 50--270           & tunable via Al content \\
    Ca$_4$Mn$_3$O$_{10}$     & $\approx$70       & Ruddlesden--Popper \\
    PYCCO                    & 60--150           & tunable via Y/Ca content \\
    BaVS$_3$                 & $\approx$82       & sulphide \\
    Fe$_3$O$_4$               & $\approx$120      & ferrimagnetic \\
    \hline
    V$_3$O$_7$                & --                & no MIT reported: the $\sim$5~K anomaly of anhydrous V$_3$O$_7$ is antiferromagnetic ordering ($T_N$) \cite{li2009_v3o7} \\
    V$_9$O$_{17}$              & $\approx$70--79   & smallest jump of the series, $\sim10^2$ \cite{kuwamoto1981_v9o17} \\
    V$_8$O$_{15}$              & $\approx$65--70   & upper edge of the HTS window \\
    V$_5$O$_9$                 & $\approx$135      & \\
    V$_2$O$_3$                 & 150--170          & only MIT compound demonstrated in-coil \\
    \hline
  \end{tabular}
\end{table}

The V--O system offers a different proposition. A sequence of
stoichiometric phases -- V$_3$O$_7$, V$_6$O$_{13}$, V$_8$O$_{15}$,
V$_9$O$_{17}$, VO$_2$ and others -- can be reached from a single
precursor chemistry by controlled oxidation or reduction. Their
reported transitions range from about 70~K to well above room
temperature (VO$_2$, $\approx$340~K). Within V$_2$O$_3$ itself,
the transition can also be shifted. Titanium doping lowers it and, above
a few percent Ti, suppresses it, while chromium doping does the opposite
and stabilises the insulating state \cite{mcwhan1973}. Oxygen excess
(V$_2$O$_{3+x}$) also lowers and finally suppresses the transition
\cite{ueda1980}. A fourth lever is strain. Epitaxial V$_x$O$_y$
nanoclusters a few nanometres in size, grown in an MgO matrix, show a
hysteretic transition as low as 1.5--27~K \cite{kengnizanguim2020_vox}.
This requires thin-film epitaxy rather than the powder and pellet
routes used here.

Some practical points must be kept in mind. Vanadium oxides can change
stoichiometry by oxidation. In our synthesis runs, however, no phase
change occurred below about 310~$^{\circ}$C (Section~\ref{sec:mat-tga}),
well above any temperature a coil hot spot is allowed to reach, and
embedding in a binder is expected to slow oxidation further. Vanadium compounds also
require careful handling: as a precaution, all of them should be
treated as potentially toxic, like V$_2$O$_5$.

This paper explores phase selection within the V--O system, mainly
toward V$_8$O$_{15}$ and V$_9$O$_{17}$, starting from the processing and
in-coil experience already gained with V$_2$O$_3$. Section~\ref{sec:materials}
describes synthesis, phase identification, specimen preparation and
measurement methods. Section~\ref{sec:electrothermal} reports the
electrical and thermal results. Section~\ref{sec:modelling} presents a
first electro-thermal coil model. Section~\ref{sec:conclusions}
summarises the results and the next steps.


\section{Materials and methods}
\label{sec:materials}

\subsection{Requirements for a smart-insulation layer}
\label{sec:mat-requirements}

An SI layer must insulate during normal operation, comparably to
conventional Kapton or fibreglass insulation. Once a hot spot heats it
above $T_{\mathrm{MIT}}$, its turn-to-turn resistance must become
comparable to, or lower than, that of a metallic turn-to-turn contact.
The primary design quantity is therefore the turn-to-turn contact
resistance $R_{\mathrm{ct}}$, or the areal resistance $R_A=\rho\,t$ of a
layer of resistivity $\rho$ and thickness $t$ \cite{ortino2024_tasc,ortino2025_tasc_mit}.
For a representative layer thickness, this corresponds to
$\rho \gtrsim 10^{2}~\Omega\,$m in the insulating state and
$\rho \approx 10^{-1}$--$10^{-6}~\Omega\,$m in the metallic state. For
REBCO magnets, $T_{\mathrm{MIT}}$ should lie between 30 and 70~K
(Section~\ref{sec:intro-strategies}).

\subsection{TGA redox synthesis}
\label{sec:mat-tga}

\begin{figure}[H]
  \centering
  \includegraphics[width=0.55\textwidth]{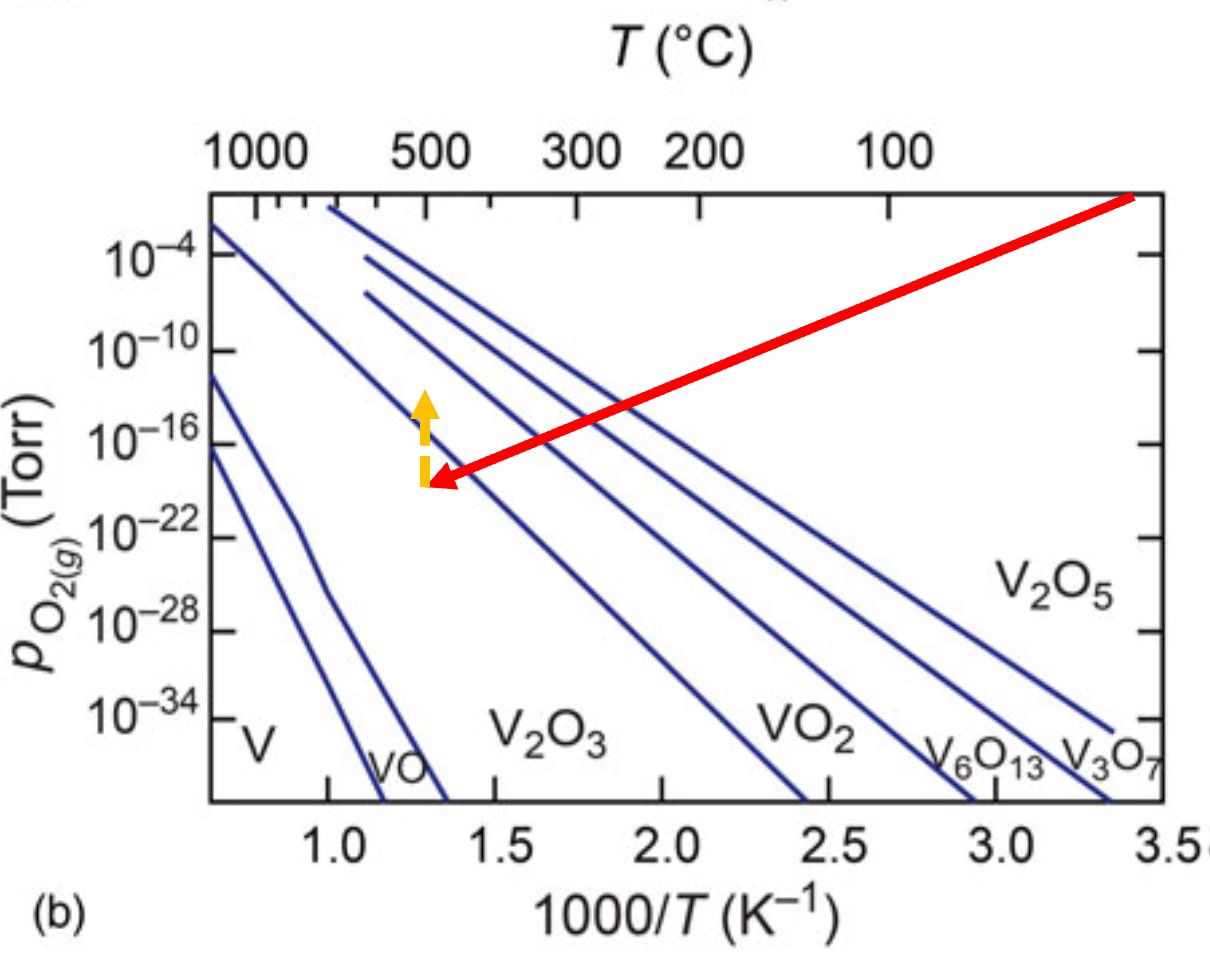}
  \caption{V--O phase diagram, adapted from \cite{hu2023}, with the
    approximate $p_{\mathrm{O}_2}$--$T$ path followed during TGA redox
    cycling (red/orange arrows), targeting the V$_2$O$_3$--VO$_2$
    boundary region.}
  \label{fig:phase-diagram}
\end{figure}

The main synthesis route uses a thermogravimetric analyser coupled to a
mass spectrometer (TGA-MS; Mettler Toledo TGA/DSC, PSI Nanopowder
Synthesis Laboratory). The precursor is either V$_2$O$_3$ (reduced) or
V$_2$O$_5$ (oxidised). Redox cycles are controlled through the oxygen
partial pressure $p_{\mathrm{O_2}}$, the temperature and the dwell time.
About twelve distinct conditions were explored in the most recent
campaigns. Two strategies were used: (i)~constant temperature, with
$p_{\mathrm{O_2}}$ tuned toward a target redox equilibrium;
(ii)~constant inlet gas composition, with a temperature sweep.
Figure~\ref{fig:phase-diagram} shows the region of the V--O phase
diagram targeted.

Reactions were active between about 310 and 700~$^{\circ}$C. No phase
change was observed below 310~$^{\circ}$C. Runs were not extended above
700~$^{\circ}$C.

\begin{figure}[H]
  \centering
  \includegraphics[width=0.5\textwidth]{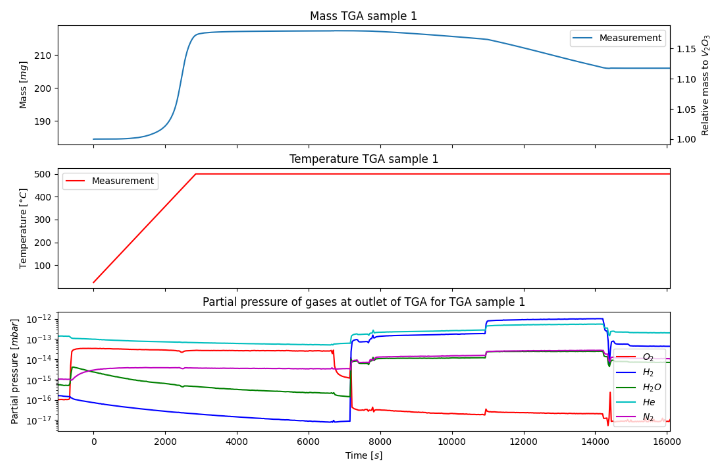}\\[2pt]
  \includegraphics[width=0.5\textwidth]{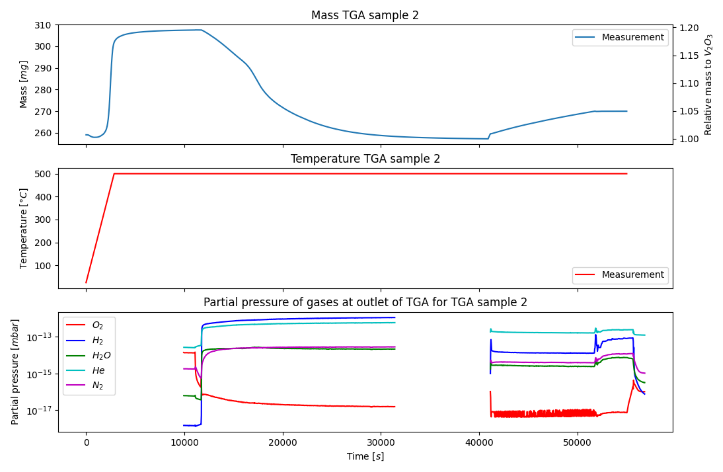}
  \caption{TGA mass and temperature, and calibrated MS partial
    pressures, for two samples synthesised with strategy (i)
    (gas-flux-controlled redox).}
  \label{fig:tga-traces}
\end{figure}

\begin{figure}[H]
  \centering
  \includegraphics[width=0.58\textwidth]{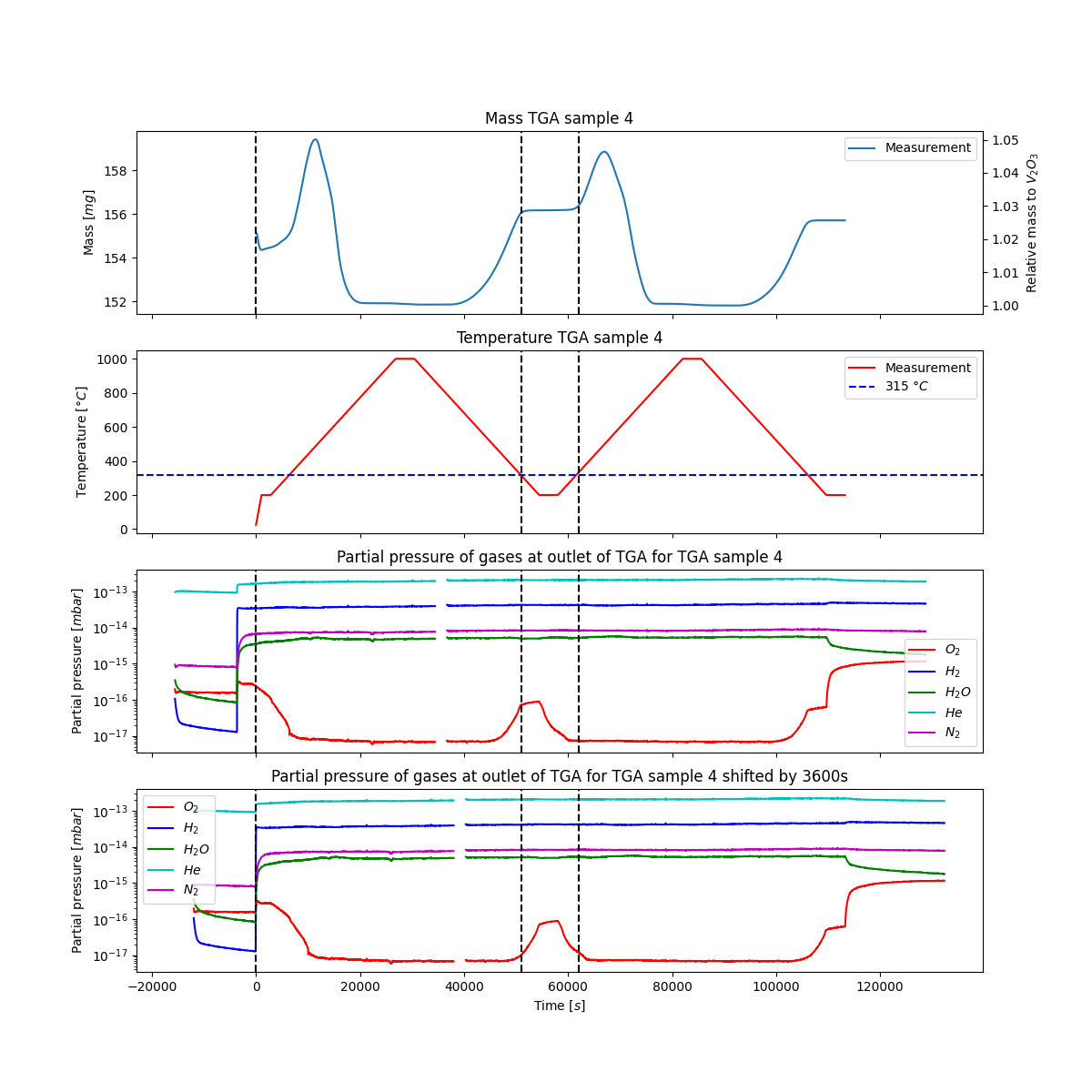}
  \caption{TGA mass and temperature, and calibrated MS partial
    pressures, for a sample synthesised with strategy (ii) (fixed inlet
    gas, temperature sweep). The reproducible mass plateaus and edges
    mark the 310--700~$^{\circ}$C range in which the redox reaction is
    active.}
  \label{fig:tga-sample4}
\end{figure}

Figure~\ref{fig:tga-traces} shows mass, temperature and outlet-gas
partial pressures for two runs using strategy (i).
Figure~\ref{fig:tga-sample4} shows a run using strategy (ii), where the
mass is steered by the temperature ramp alone. The mass-spectrometer
ion currents were converted into partial pressures through a
calibration.

Repeated runs showed reproducible mass plateaus and edges, so the redox
state can be steered with good mass resolution. A stable mass plateau,
however, does not guarantee complete conversion to a single target
phase. Two explanations are possible. Conversion may be incomplete
within each particle, with a reacted surface and a less converted
core. Or it may be inhomogeneous across the powder bed in the crucible,
for example through gas-access or temperature gradients. The two cases
call for opposite remedies: smaller particles would help in the first,
whereas a thinner or mixed bed would help in the second. The present
data do not distinguish between them.

\subsection{Phase identification by XRPD}
\label{sec:mat-xrpd}

\begin{figure}[tbp]
  \centering
  \begin{tabular}{cc}
    \includegraphics[width=0.47\textwidth]{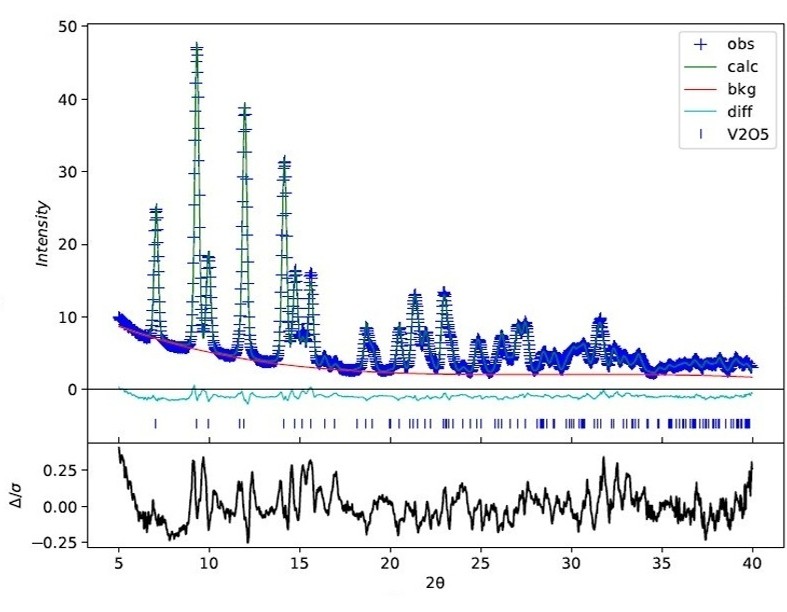} &
    \includegraphics[width=0.47\textwidth]{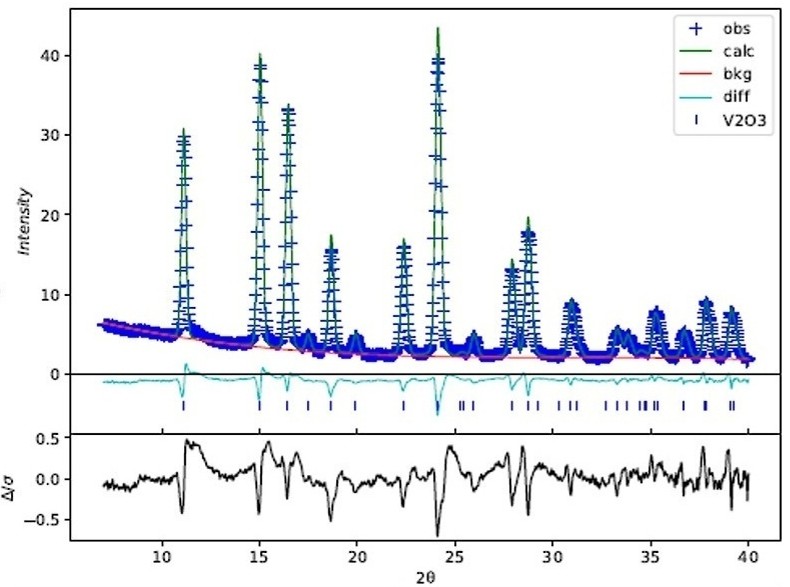} \\
    (a) & (b) \\[0.8em]
    \includegraphics[width=0.47\textwidth]{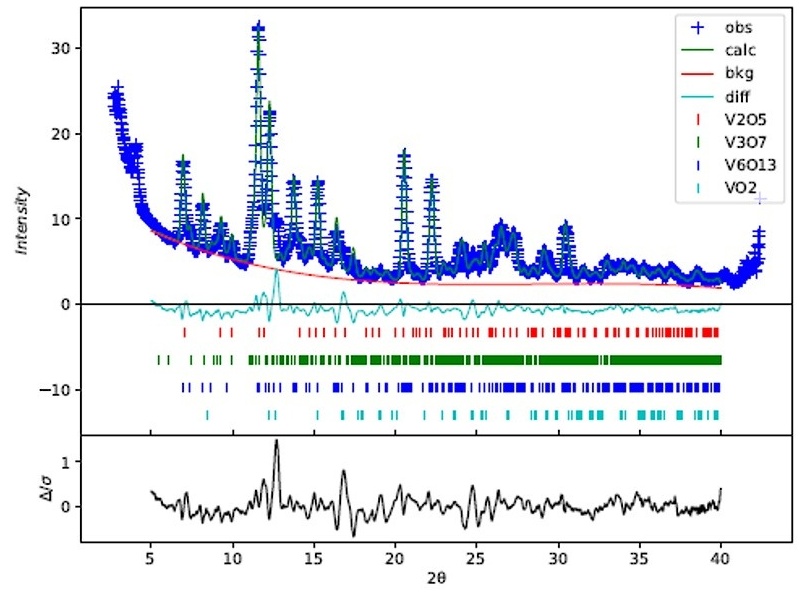} &
    \includegraphics[width=0.47\textwidth]{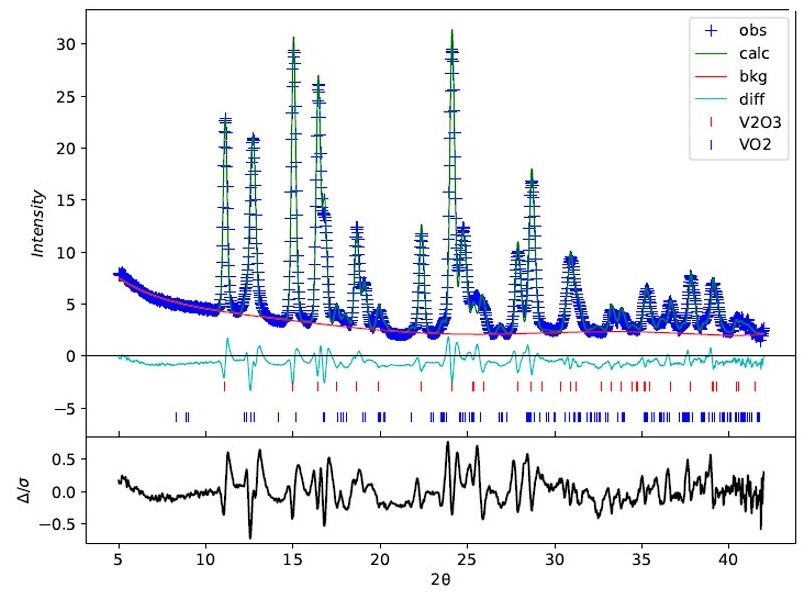} \\
    (c) & (d) \\
  \end{tabular}
  \caption{Representative XRPD patterns and Rietveld refinements of
    precursors (V$_2$O$_3$, V$_2$O$_5$) and TGA-processed products,
    showing typical multi-phase outcomes.}
  \label{fig:xrpd-rietveld}
\end{figure}

Phases were identified by X-ray powder diffraction (XRPD) at PSI. A
laboratory diffractometer (Bruker) was used for fast screening. Higher
accuracy patterns were measured at the ADDAMS end-station of the Swiss
Light Source \cite{willmott2013_addams}, with powders loaded into
0.3--0.5~mm silica capillaries in transmission geometry. Phase fractions
were obtained by Rietveld refinement in GSAS-II
(Fig.~\ref{fig:xrpd-rietveld}). The TGA route consistently gave
multi-phase powders. Reduced target phases, such as V$_3$O$_7$ or
V$_8$O$_{15}$, typically appeared as minor fractions alongside V$_2$O$_3$,
V$_2$O$_5$, V$_6$O$_{13}$ and VO$_2$. These refinements were used to
select the batches carried forward to electrical testing.

\subsection{Hydrothermal synthesis of V$_3$O$_7\cdot$H$_2$O}
\label{sec:mat-hydrothermal}

The TGA route gave V$_3$O$_7$ only as a minor fraction. A hydrothermal
route adapted from \cite{chine2011} was therefore tested to obtain it
directly. Oxalic acid (5.2~g) was stirred in 50~mL of deionised water,
and V$_2$O$_5$ (2.5~g) was added. The suspension was stirred for 2~h at
room temperature and for about 5~h at 80~$^{\circ}$C, giving a dark-blue
solution. This was dried at 100~$^{\circ}$C for 16~h and the resulting
powder was calcined at 400~$^{\circ}$C for 10~h (2~K/min), giving an
orange powder. About 1~g of this powder was refluxed at 110~$^{\circ}$C
for 16~h in 10~mL of 0.025~M aqueous ascorbic acid, diluted to 20~mL
with water. The green,
gel-like suspension was then treated in a 50~mL PTFE-lined autoclave
at 220~$^{\circ}$C for 6~h. The green precipitate was filtered, washed
with water and isopropanol, and dried at 80~$^{\circ}$C for 7~h.
Reproducibility and scale-up were
limited by incomplete dissolution of oxalic acid at room temperature, a
gel-like suspension that complicated transfer, and slow vacuum
filtration.

\begin{figure}[H]
  \centering
  \begin{minipage}{0.38\textwidth}
    \centering
    \includegraphics[width=\textwidth]{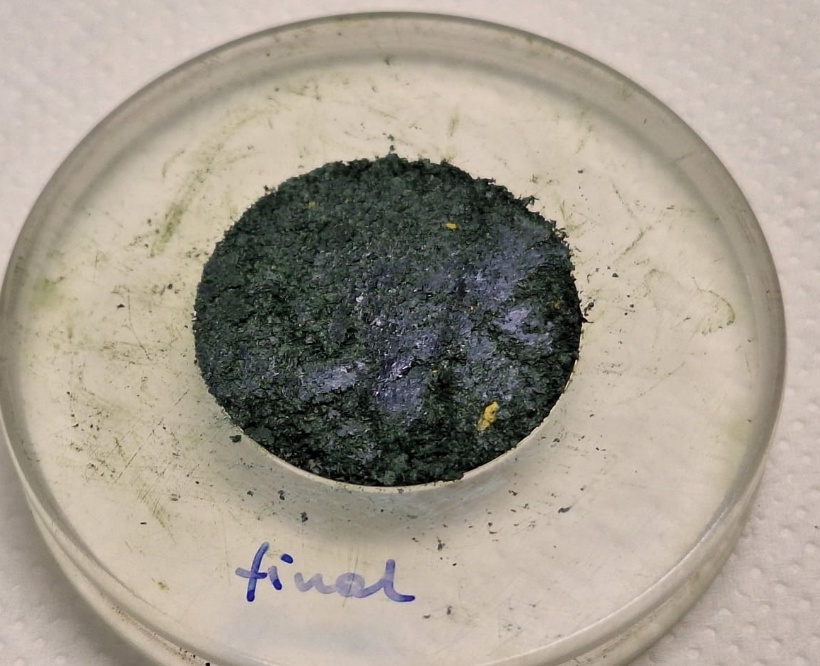}\\
    (a)
  \end{minipage}\hfill
  \begin{minipage}{0.52\textwidth}
    \centering
    \includegraphics[width=\textwidth]{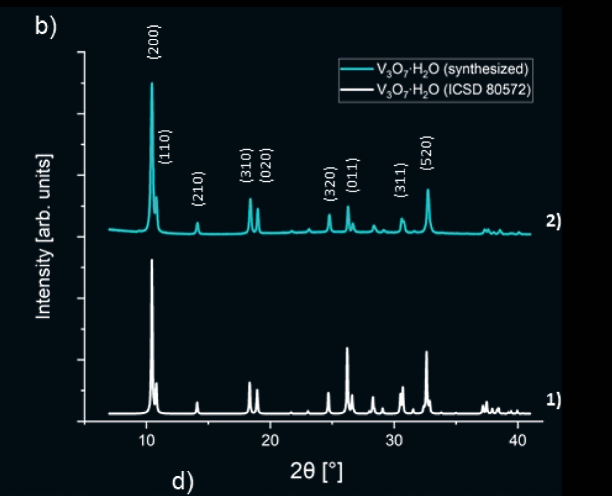}\\
    (b)
  \end{minipage}
  \caption{Hydrothermal product: (a) dark-green V$_3$O$_7\cdot$H$_2$O
    powder; (b) XRPD pattern compared with the reference pattern
    ICSD~80572.}
  \label{fig:hydrothermal-xrpd}
\end{figure}

\begin{figure}[H]
  \centering
  \begin{tabular}{ccc}
    \includegraphics[width=0.30\textwidth]{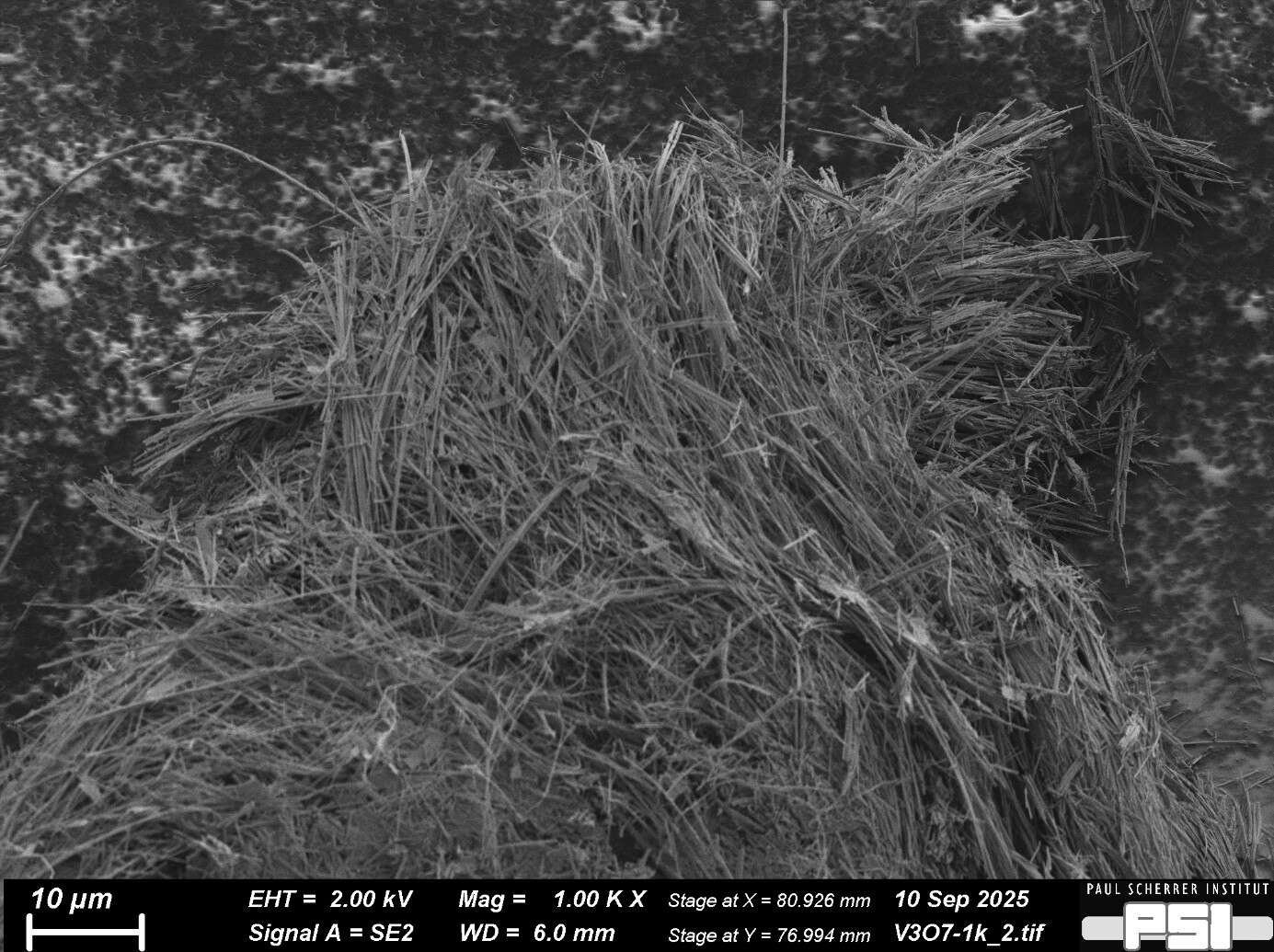} &
    \includegraphics[width=0.30\textwidth]{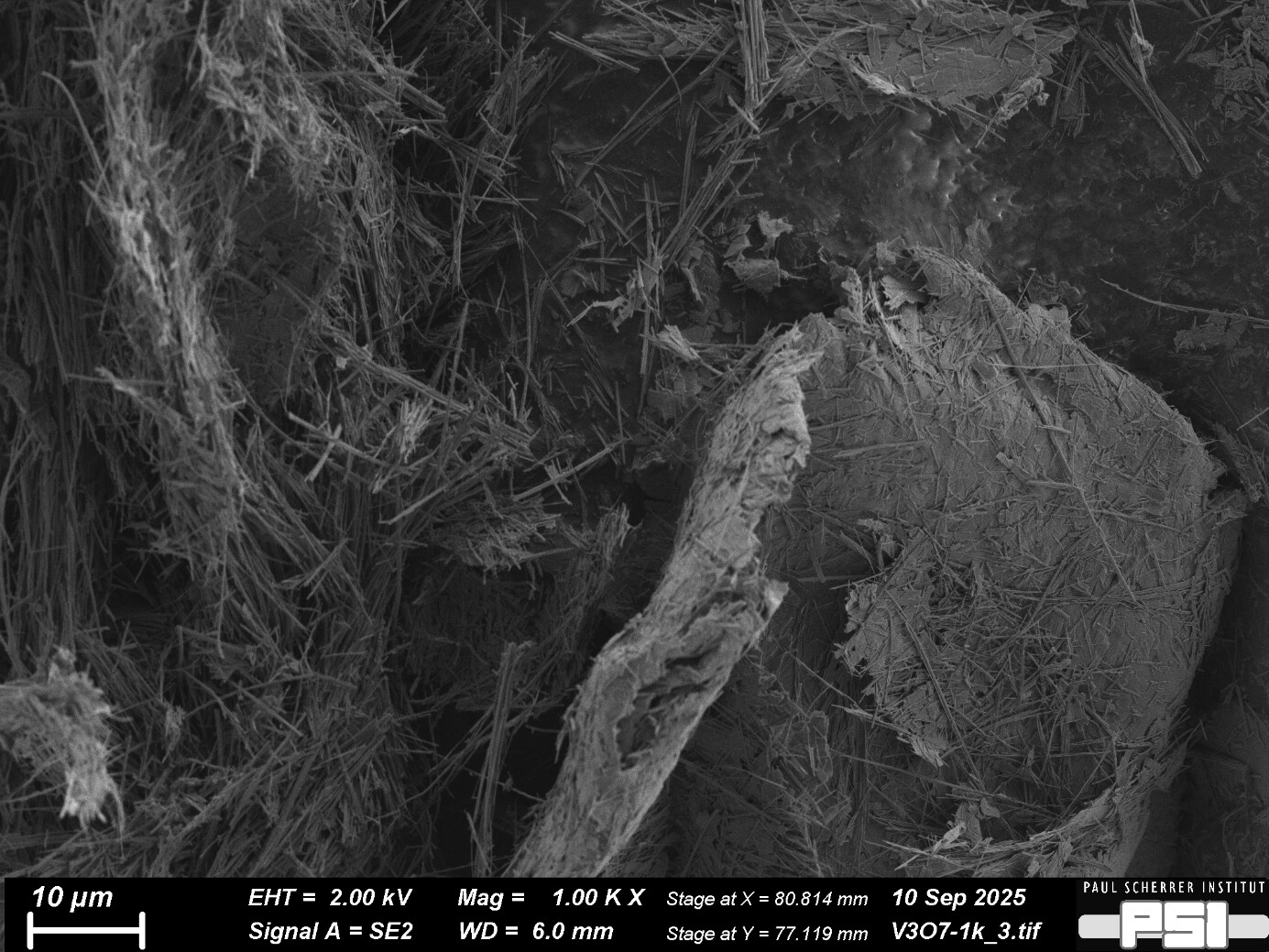} &
    \includegraphics[width=0.30\textwidth]{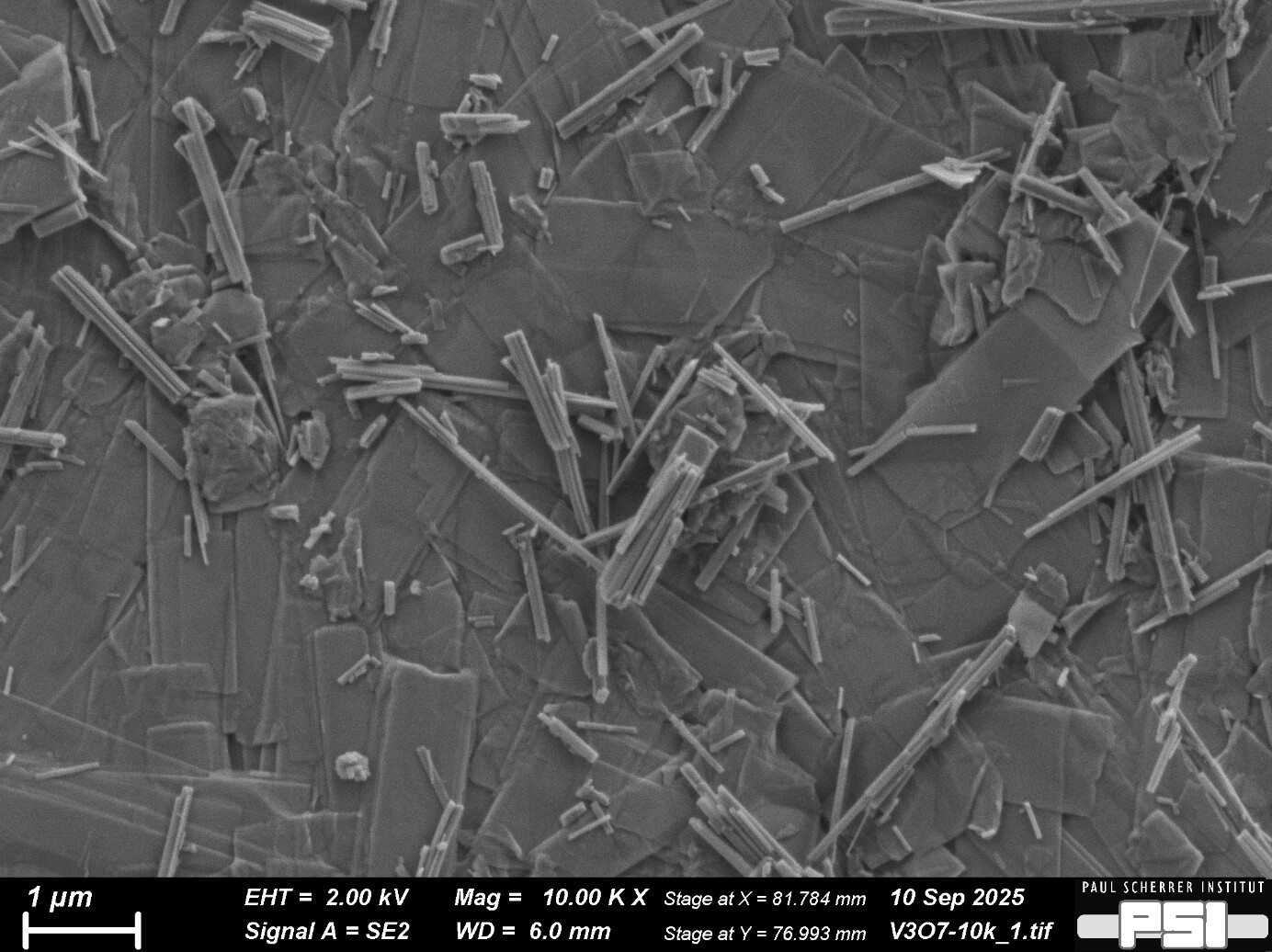} \\
    (a) & (b) & (c) \\
  \end{tabular}
  \caption{SEM micrographs of the hydrothermal product: (a,b)
    1000$\times$, bundles of belt- and needle-like crystallites;
    (c) 10\,000$\times$, coexisting platelet and rod habits.}
  \label{fig:sem-v3o7}
\end{figure}

The product crystallised as the hydrate V$_3$O$_7\cdot$H$_2$O
(Fig.~\ref{fig:hydrothermal-xrpd}), whose XRPD pattern matches the
reference pattern ICSD~80572 in peak positions. This is an important
difference from the literature on the low-temperature anomaly of
V$_3$O$_7$, which concerns the anhydrous phase \cite{li2009_v3o7}.
Part of the product was therefore annealed at 400 and 500~$^{\circ}$C
to remove the structural water before electrical testing
(Section~\ref{sec:et-v3o7}). Scanning electron microscopy (SEM,
Fig.~\ref{fig:sem-v3o7}) shows dense bundles of thin belt- or
needle-like crystallites at 1000$\times$. At 10\,000$\times$, two crystal
habits coexist in the same batch: compact platelets with polygonal
facets, and thinner rods lying across them.

XRPD confirms that annealing at 400 and 500~$^{\circ}$C removed the
structural water: the annealed material is anhydrous V$_3$O$_7$, with
V$_2$O$_5$ as a secondary phase from partial oxidation.

\subsection{Specimens for electrical testing}
\label{sec:mat-processing}

\begin{figure}[H]
  \centering
  \begin{tabular}{ccc}
    \raisebox{0.6cm}{\includegraphics[width=0.42\textwidth]{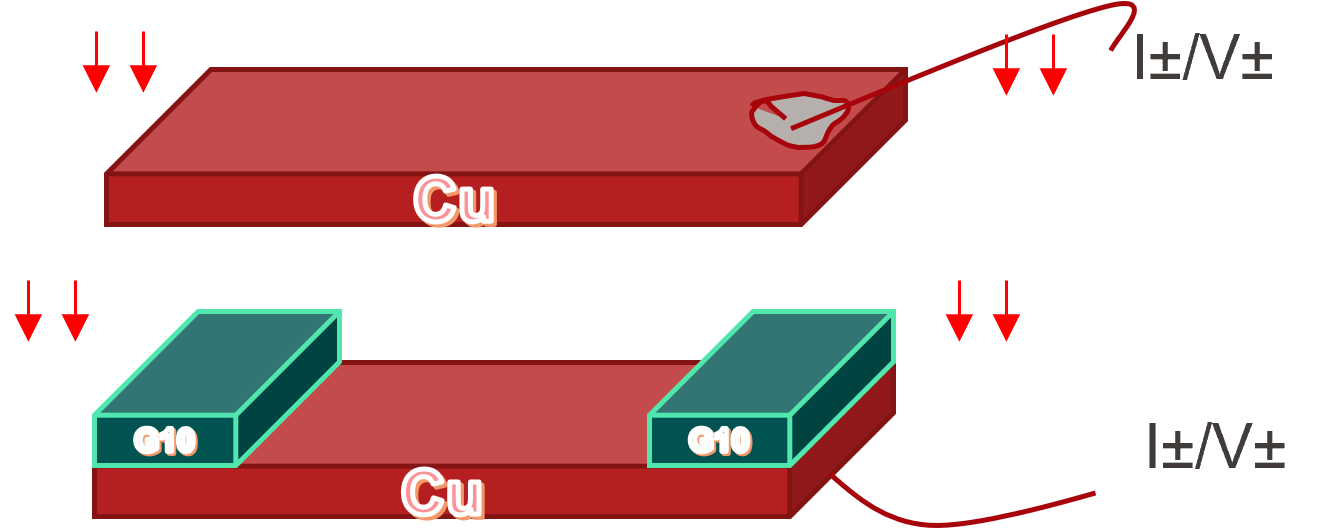}} &
    \includegraphics[height=5.2cm]{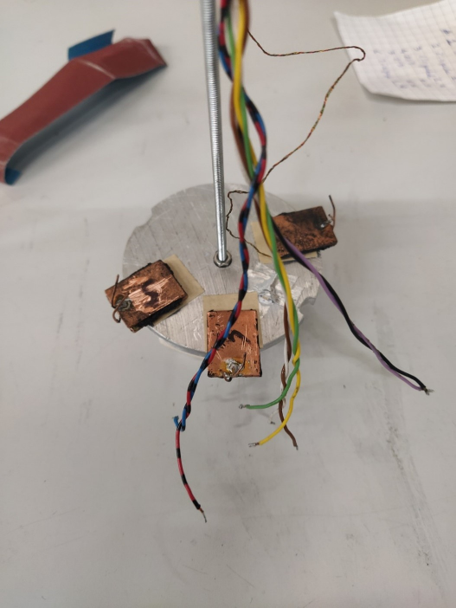} &
    \includegraphics[height=5.2cm]{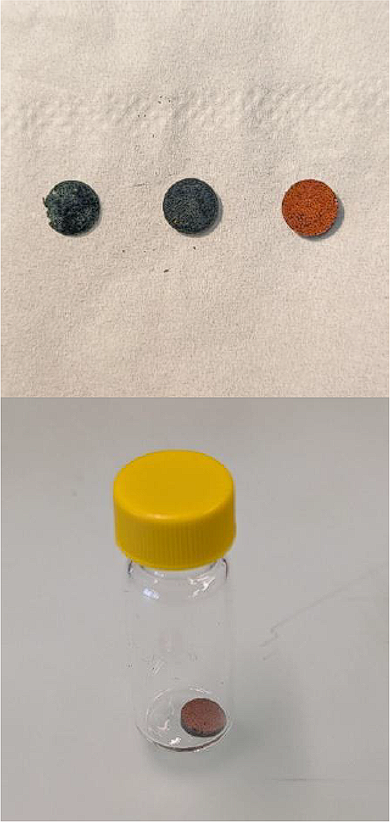} \\
    (a) & (b) & (c) \\
  \end{tabular}
  \caption{Specimen types: (a) epoxy-bound Cu-sandwich geometry, with
    two-terminal contacts on the Cu plates (top) and four-terminal,
    G10-spaced contacts (bottom); (b) V$_2$O$_3$--epoxy Cu sandwiches
    wired on the aluminium holder of the bench-top set-up; (c)
    binder-free pressed pellets (15~mm discs), and a pellet with
    Cu-microsphere electrode layers in a vial.}
  \label{fig:specimens}
\end{figure}

Two specimen types were used (Fig.~\ref{fig:specimens}).

\textbf{Epoxy-bound Cu sandwiches.} Commercial V$_2$O$_3$ (95\% purity)
was used to develop the processing route. Powder and epoxy were mixed
and cast between two Cu electrodes (contact area $20\times5$~mm). A
0.4~mm G10 spacer set the layer thickness. The electrodes were clamped
during curing. Current flows through the layer thickness, between the
two Cu plates. Several binders were screened
(Table~\ref{tab:binder-comparison}): Araldite\textregistered\
CY221/Aradur~2967 (the standard SULTAN-sample epoxy), a
methyl-methacrylate epoxy, Stycast~2850, and two conductive graphite
coatings. Cyanate-ester systems (CTD~101/400/403) remain to be tested.
Phenyldodecane and Evonik EPODIL~750 were used as diluents to control
viscosity. Across about 25 samples, a V$_2$O$_3$:epoxy weight ratio of
1.5:1 performed best. The curing atmosphere (air, vacuum or oven) had no
major effect.

\textbf{Binder-free pressed pellets.} Epoxy raises the warm-state
resistivity by about two to three orders of magnitude compared with the
bare powder. Binder-free pellets were therefore also prepared. Powders
were pressed at 2~t into 15~mm discs, 0.1--0.5~mm thick, either alone or
with Cu-microsphere layers pressed on both faces as electrodes. Both
V$_2$O$_3$ and V$_3$O$_7\cdot$H$_2$O compacted well; the hydrate was
easier to compact. The applied pressure is far below the range where
pressure-induced phase changes would matter. The pellets showed no
degradation over time.

Table~\ref{tab:binder-comparison} compares the room-temperature
resistivity of each binder system with the insulating target. Where the
geometry does not allow a clean conversion, the raw resistance is
given. Earlier Stycast~2850 samples (50--52), whose electrodes were
embedded in the resin before curing, are excluded: their very high
resistance could not be distinguished from a contact artefact. Sample
53, with verified electrode contact, is reported instead.

\begin{table}[H]
\centering
\caption{Room-temperature resistivity (or resistance, where the
  geometry does not allow a $\rho$ conversion) of V$_2$O$_3$ composite
  specimens with different binders, compared with the
  $\rho\gtrsim10^2~\Omega\,$m insulating target.}
\label{tab:binder-comparison}
\begin{tabular}{llp{5.2cm}}
\hline
Binder & $\rho_{RT}$ or $R_{RT}$ & Note \\
\hline
Araldite\textregistered\ CY221/Aradur~2967 & $\rho_{RT}\sim10^{-1}$~$\Omega\,$m & standard SULTAN epoxy; best recipe \\
Methyl-methacrylate epoxy & $R_{RT}\gtrsim10^{4}~\Omega$ & similar $\rho$; good cryogenic behaviour \\
Stycast~2850 & $\rho_{RT}\approx2.4\times10^{4}~\Omega\,$m & well above target; verified contact \\
838-C graphite coating & $R_{RT}\approx1.3~\Omega$ & far too conductive \\
33-Graphite coating & $R_{RT}\approx0.3~\Omega$ & far too conductive \\
\hline
\end{tabular}
\end{table}

One processing result was counter-intuitive. Grinding the V$_2$O$_3$
powder 3--5 times finer \emph{raised} the composite resistance. Coarser
grains gave lower, i.e.\ better, metallic-state resistance.

\subsection{Electrical and thermal measurements}
\label{sec:mat-measurements}

$R(T)$ was measured in three set-ups of increasing cryogenic reach.

\textbf{Bench-top, 77--300~K.} Wired specimens on an aluminium holder
were lowered slowly into a liquid-nitrogen dewar. A constant-current
source and multi-channel voltmeters logged $R(T)$. This set-up was used
for fast screening.

\begin{figure}[tbp]
  \centering
  \includegraphics[width=0.3\textwidth]{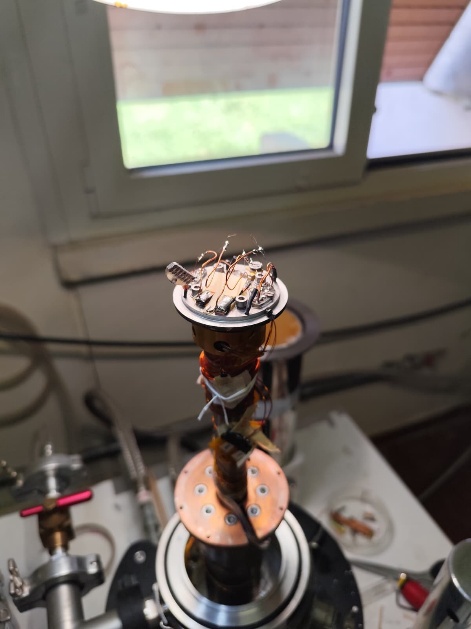}
  \caption{ENEA-Frascati cryobox: cold finger with mounted Cu-sandwich
    specimens.}
  \label{fig:enea-setup}
\end{figure}

\textbf{ENEA-Frascati cryobox, down to $\approx$3~K.} Cu-sandwich
specimens were measured in a conduction-cooled cryostat (September and
November 2025; Fig.~\ref{fig:enea-setup}). A magnetic field up to 2~T
can be applied. The logged sweep rate allowed each data point to be
assigned to a warming or a cooling leg.

\textbf{PPMS, down to 4.2~K.} Binder-free V$_3$O$_7\cdot$H$_2$O pellets
were measured at PSI with the resistivity option of a Physical Property
Measurement System (PPMS). Two pellets were 15~mm discs, 0.3~mm thick
(as-synthesised and annealed at 400~$^{\circ}$C). The third was a small
strip about 0.2~mm thick (annealed at 500~$^{\circ}$C). Four wires were
bonded on the top face of each specimen. The measurement is therefore
an in-plane four-point measurement, not a through-thickness one as for
the Cu sandwiches. Temperature resolution was 10--100 points per kelvin.

\textbf{Specific heat.} $c_p(T)$ was measured with the PPMS heat-capacity
option (He4 puck 3261) on a pressed V$_2$O$_3$ pellet (95\% purity,
28.8~mg, Apiezon~N grease), from 300 to 4~K at zero field.


\section{Electrical and thermal results}
\label{sec:electrothermal}

\subsection{V$_2$O$_3$--epoxy reference and acceptance criterion}
\label{sec:et-criterion}

\begin{figure}[H]
  \centering
  \includegraphics[width=0.75\textwidth]{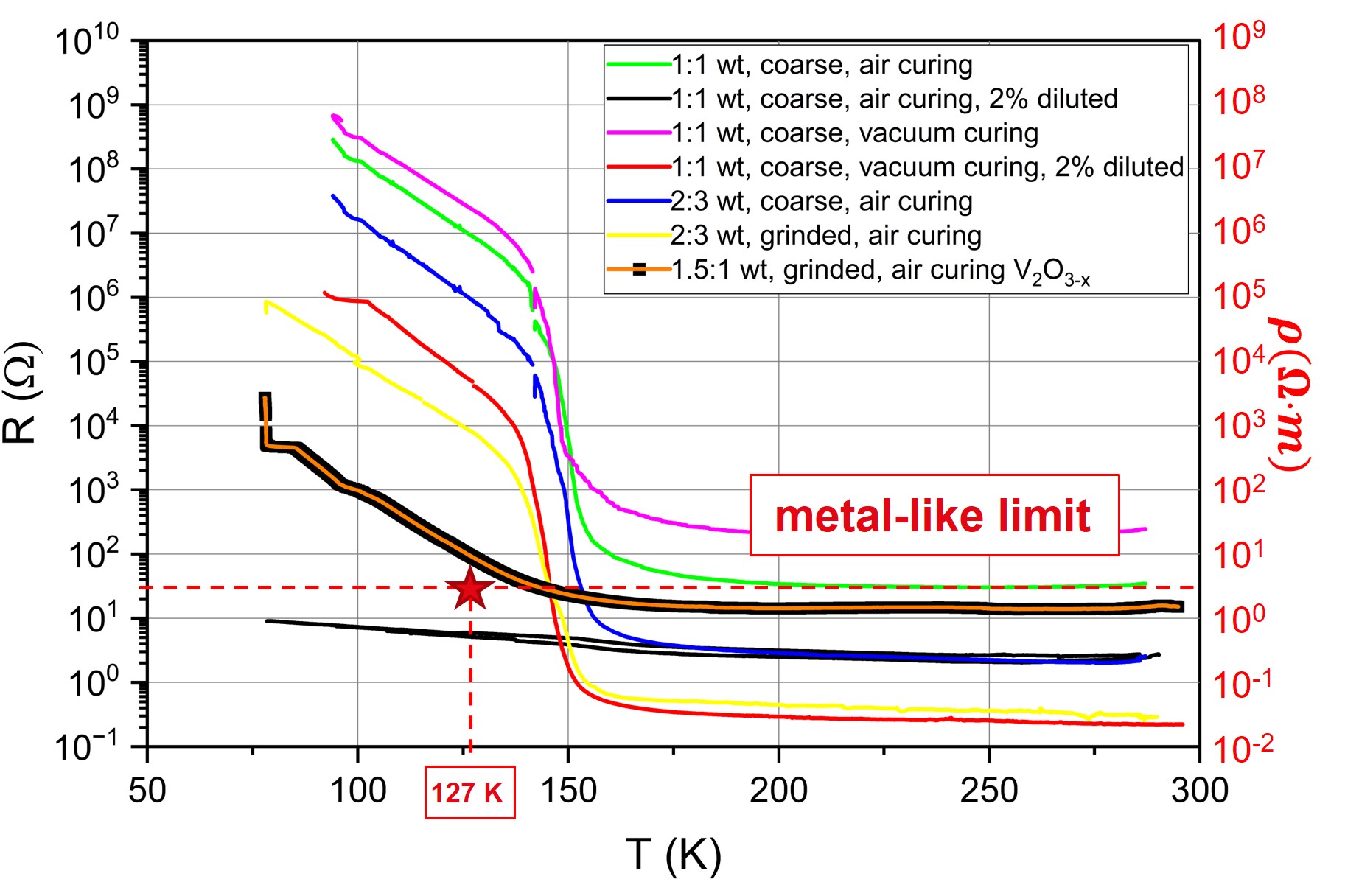}
  \caption{Resistance of V$_2$O$_3$--epoxy composites between room
    temperature and 77~K, across processing variants, defining the
    metal-like acceptance limit used to down-select new V$_x$O$_y$
    batches.}
  \label{fig:v2o3-rt}
\end{figure}

Figure~\ref{fig:v2o3-rt} shows the room-temperature-to-77~K resistance
of V$_2$O$_3$--epoxy specimens across the processing variants of
Section~\ref{sec:mat-processing}. The transition temperature did not
depend on the binder chemistry. No degradation was observed after
repeated cycling between 77~K and room temperature. Background fields
up to 15~T, tested down to 4.2~K, had no measurable effect on the
insulating state below the transition. From the best specimens we
defined a practical acceptance criterion, the ``metal-like limit''.
It is the room-temperature resistance that a new powder must reach,
in the same specimen geometry, to be considered further.

\subsection{Multi-phase TGA powders}
\label{sec:et-tga}

The first phase-modified TGA batch contained 43\% V$_6$O$_{13}$, 35\%
VO$_2$, 11\% V$_3$O$_7$ and 9\% V$_2$O$_3$ (Rietveld). Its effective
transition temperature was $\approx$127~K, about 25--28~K below that of
V$_2$O$_3$. This is a first proof that changing the oxide
stoichiometry moves the transition in the intended direction. About
20--25 further samples confirmed the trend. In the most recent batches,
the V$_8$O$_{15}$ fraction was raised from $\approx$42\% to above 50\%
by tuning the synthesis parameters alone.

\begin{figure}[H]
  \centering
  \includegraphics[width=0.8\textwidth]{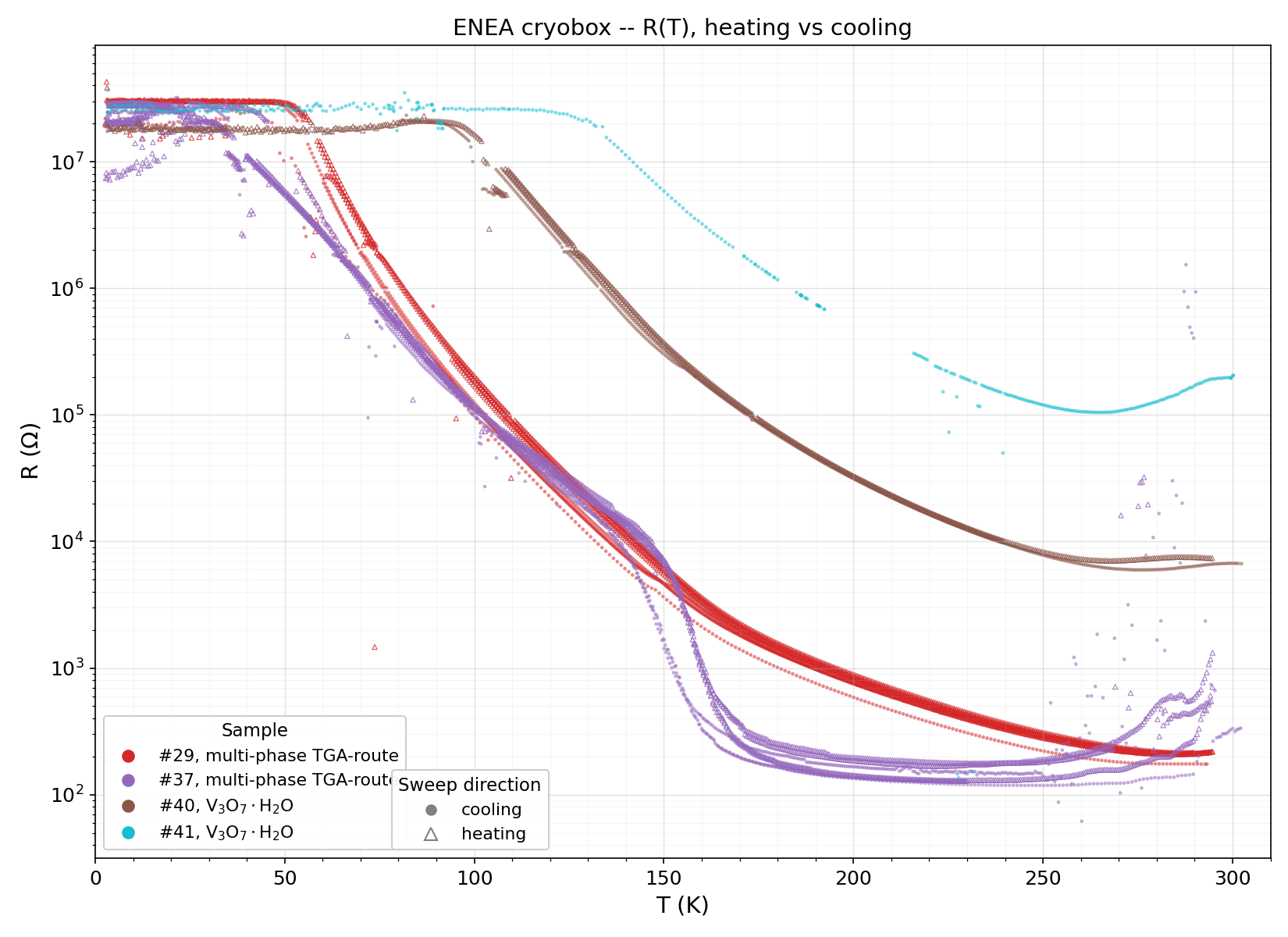}
  \caption{ENEA-Frascati $R(T)$ of epoxy-bound Cu-sandwich specimens.
    Samples \#29 and \#37: multi-phase TGA powders. Samples \#40 and
    \#41: V$_3$O$_7\cdot$H$_2$O. Filled circles: cooling; open
    triangles: heating, assigned point by point from the measured
    $\mathrm{d}T/\mathrm{d}t$. Sample \#41 was measured on cooling
    only.}
  \label{fig:enea-rt}
\end{figure}

Figure~\ref{fig:enea-rt} shows the ENEA results. Samples \#29 and \#37
are multi-phase TGA powders, of the kind shown in
Fig.~\ref{fig:xrpd-rietveld}c. Both show a resistance drop on warming,
but a broad one, spread over several tens of kelvin. This is consistent
with minority MIT-active phases rather than with one sharp transition.

All samples measured in both directions (\#29, \#37, \#40) show the
same splitting: $R$ is higher on warming than on cooling over
$\sim$140--250~K. A similar splitting had been seen in our bench-top
measurements down to 77~K. There, it could be explained by the distance
between the Cernox thermometer and the specimen: the sensor cools or
warms before the specimen as the holder enters or leaves the
liquid-nitrogen bath. No such offset exists in the ENEA cryobox, so this explanation does
not apply. The loop is therefore more likely intrinsic, i.e.\ a thermal
hysteresis such as that of first-order transitions in V$_2$O$_3$
\cite{kim2019_racetrack}.

\subsection{Hydrothermal V$_3$O$_7\cdot$H$_2$O}
\label{sec:et-v3o7}

\begin{figure}[H]
  \centering
  \includegraphics[width=0.62\textwidth]{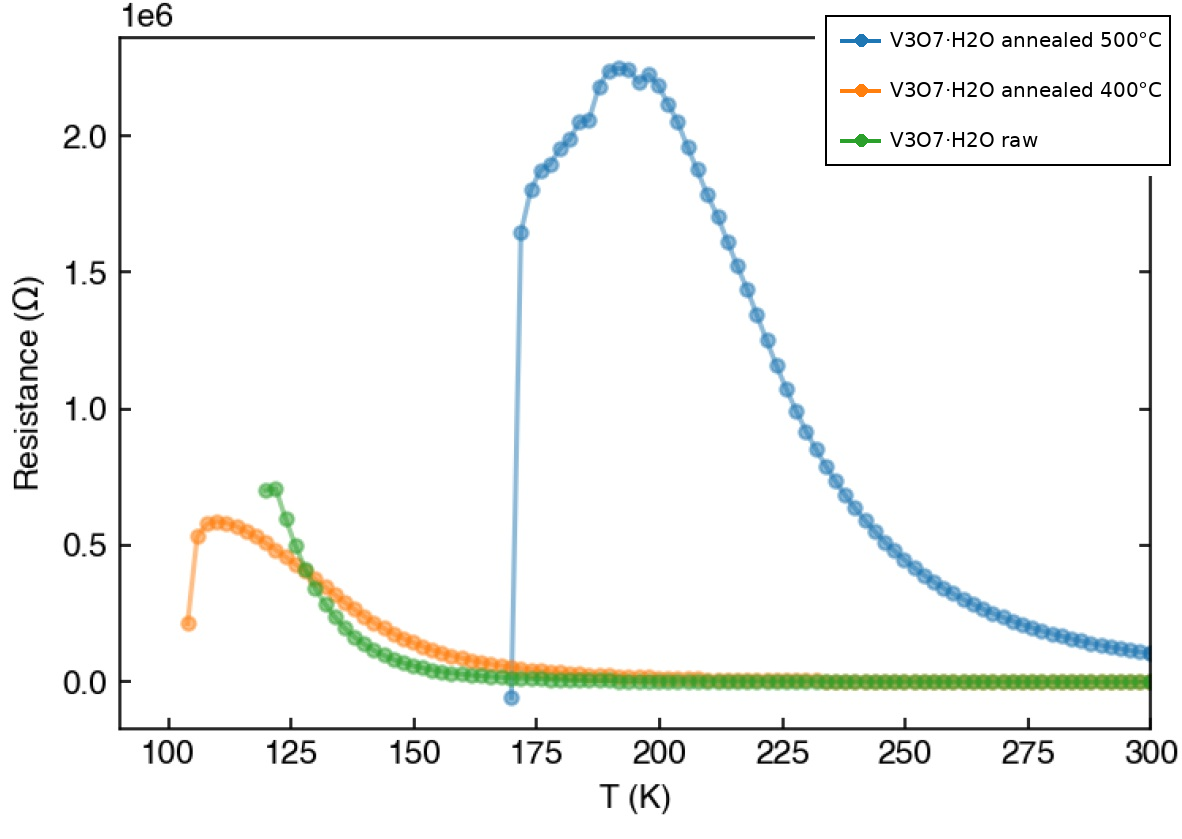}
  \caption{PPMS four-point $R(T)$ of bare V$_3$O$_7\cdot$H$_2$O pellets:
    as-synthesised, and annealed at 400 and 500~$^{\circ}$C. The
    500~$^{\circ}$C run is cut off near 170~K by the compliance voltage
    of the current source.}
  \label{fig:ppms-rt}
\end{figure}

V$_3$O$_7\cdot$H$_2$O was measured in two forms: epoxy-bound in the
ENEA campaign (\#40, \#41; Fig.~\ref{fig:enea-rt}) and as bare pellets
in the PPMS (as-synthesised, and annealed at 400 and 500~$^{\circ}$C;
Fig.~\ref{fig:ppms-rt}). All specimens were strongly resistive. Above
$\sim$100~K their resistance was one to three orders of magnitude higher
than that of the multi-phase TGA samples, and it rose steadily on
cooling without a sharp step. The bare pellets already showed hundreds of
k$\Omega$ at room temperature. Several PPMS runs reached the compliance
voltage of the current source before any low-temperature feature could
be resolved. The 500~$^{\circ}$C run, for example, stops near 170~K. The
practical conclusion is direct. An SI layer needs a low resistance
\emph{above} its transition. Even if this material had a transition
near 5~K, it would be far too resistive above it to be useful.

The literature, examined after these measurements, points the same
way. Reports of a sharp anomaly near 5~K in V$_3$O$_7$ are scarce. NMR
and neutron-scattering studies on anhydrous V$_3$O$_7$ identify it as
the N\'eel temperature of an antiferromagnetic ordering, with no
structural or metal--insulator transition down to 4.2~K
\cite{li2009_v3o7}. The few reports of a $\sim$5~K feature most
probably concern this magnetic transition. Our as-synthesised product was the
hydrate. XRPD after annealing at 400 and 500~$^{\circ}$C shows that it
converted to anhydrous V$_3$O$_7$, with V$_2$O$_5$ as a secondary phase
(Section~\ref{sec:mat-hydrothermal}). The annealed pellets were still
far too resistive and showed no transition. Either way, V$_3$O$_7$ does
not offer a usable MIT, and this
route was closed. The low-temperature target moved to V$_8$O$_{15}$
($T_{\mathrm{MIT}}\approx65$--$70$~K), which lies at the upper edge of
the HTS window.

\subsection{Specific heat of V$_2$O$_3$}
\label{sec:et-cp}

\begin{figure}[H]
  \centering
  \begin{minipage}{0.45\textwidth}
    \centering \includegraphics[width=\textwidth]{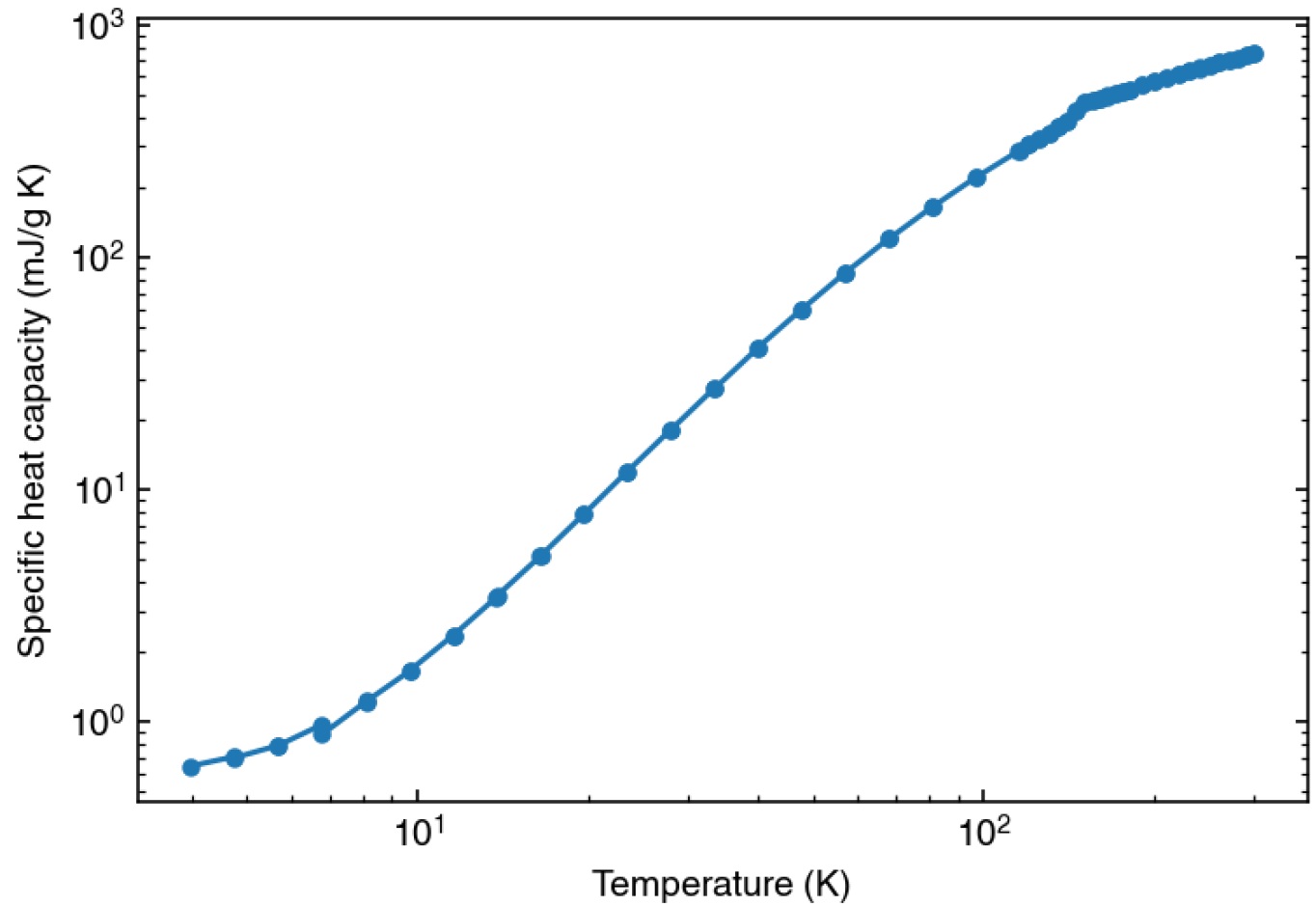}\\ (a)
  \end{minipage}\hfill
  \begin{minipage}{0.45\textwidth}
    \centering \includegraphics[width=\textwidth]{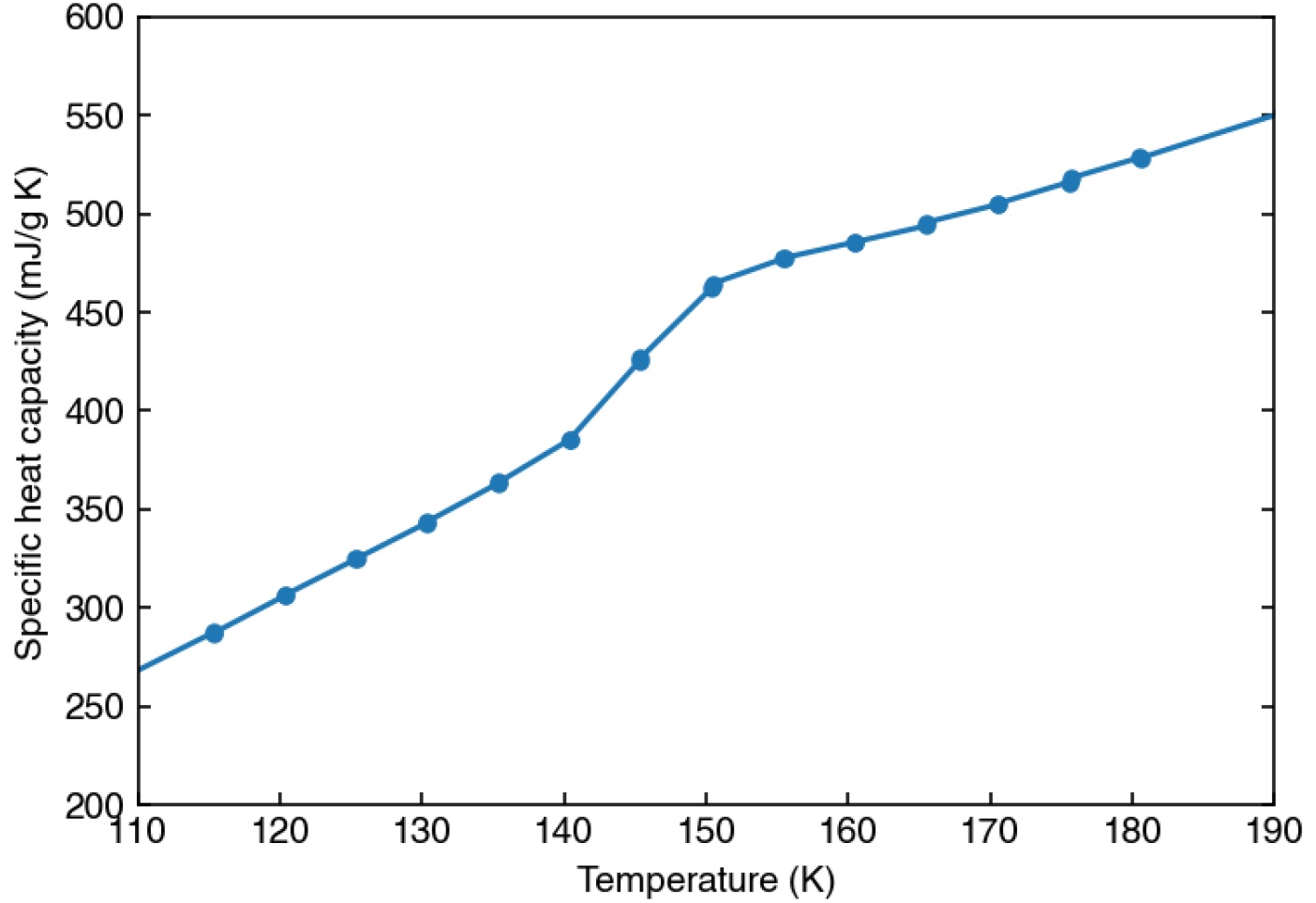}\\ (b)
  \end{minipage}
  \caption{Specific heat $c_p(T)$ of a pressed V$_2$O$_3$ pellet (PPMS,
    zero field, 300--4~K): (a) full range, log--log; (b) linear detail
    near the transition.}
  \label{fig:cp-v2o3}
\end{figure}

Figure~\ref{fig:cp-v2o3} shows $c_p(T)$ of pressed V$_2$O$_3$. The
thermal coupling between sample and platform stayed above 95\% above
the transition and above 85\% below it. It showed a clear discontinuity
exactly at the transition. This points to a significant electronic
contribution to the thermal conductivity there, and to a marked
first-order character. The $c_p$ anomaly at the transition is modest,
about 5\% of the total. Below $T_{\mathrm{MIT}}$, $c_p$ follows an
approximately $T^3$ phonon law, with a lower effective exponent below
$\approx$10~K. No further anomaly appears down to 4~K. These data
replace the literature-based $c_p(T)$ used in our first quench model
(Section~\ref{sec:modelling}), which differed from the measurement by
close to two orders of magnitude.

\section{Electro-thermal modelling: first results}
\label{sec:modelling}

The requirements of Section~\ref{sec:mat-requirements} come from
coil-scale modelling. A 3D lumped-element model of an NI coil was
developed in MATLAB \cite{ortino2024_tasc,ortino2025_tasc_mit}. In that
model, the turn-to-turn contact resistance $R_{\mathrm{ct}}$ governs
current redistribution and heat spreading during a quench. A 2D COMSOL
model was then built to include a smart-insulation layer explicitly and
to use measured material properties. This section gives an outline and
a first reference case. The full coil-scale study will be reported
separately.

\subsection{Model}
\label{sec:model-electrothermal}

The model couples three physics interfaces. An electrical circuit
represents the coil as an inductor carrying an initial current $I_0$.
An electric-current interface solves the current density in the
conductor and in the SI layer. A heat-transfer interface is coupled to
it through the Ohmic heat source. The quench is triggered by a short
Gaussian heat pulse at a chosen point of the winding. Coil radius, turn
count and SI-layer thickness are parameters.

The SI layer was first described by a ``virtual SI'' material:
literature $c_p(T)$ \cite{surikov2021} and $k(T)$ \cite{andreev1978} of
V$_2$O$_3$, with the transition shifted to 25~K and mixed 1:1 by mass
with epoxy. The measured $c_p(T)$ of Section~\ref{sec:et-cp} differs from
this input by close to two orders of magnitude. Propagating the
measured data into the model is the next step.

Two numerical measures were necessary. The resistivity steps of the
superconductor (at $T_c$) and of the SI layer (at $T_{\mathrm{MIT}}$)
had to be smoothed, otherwise the Newton iteration did not converge.
The time step had to be fixed at about 1~ms, because automatic
time-stepping skipped steps during the fast transient and violated
energy conservation.

\subsection{Reference case}
\label{sec:model-results}

The reference case is a 3-turn coil with an inner radius of 5~cm,
$I_0 = 1$~kA and a lumped inductance $L = 80~\mu$H, i.e.\ a stored
energy of 40~J. The superconductor has $T_c = 20$~K, the virtual SI
layer $T_{\mathrm{MIT}} = 25$~K, and the initial temperature is 12~K.
An 8~J Gaussian pulse, centred at $t = 0.5$~s and lasting 0.1~s,
triggers the quench on the innermost turn. The pulse is longer, and its
energy higher, than a real trigger. This was chosen for numerical
stability. Note that $T_{\mathrm{MIT}} = 25$~K was set before the
30--70~K window was defined.
The inductance study gives $L\approx1.3~\mu$H for this geometry;
the lumped value of 80~$\mu$H was set by hand to raise the stored
energy.

\begin{figure}[tbp]
  \centering
  \includegraphics[width=0.85\textwidth]{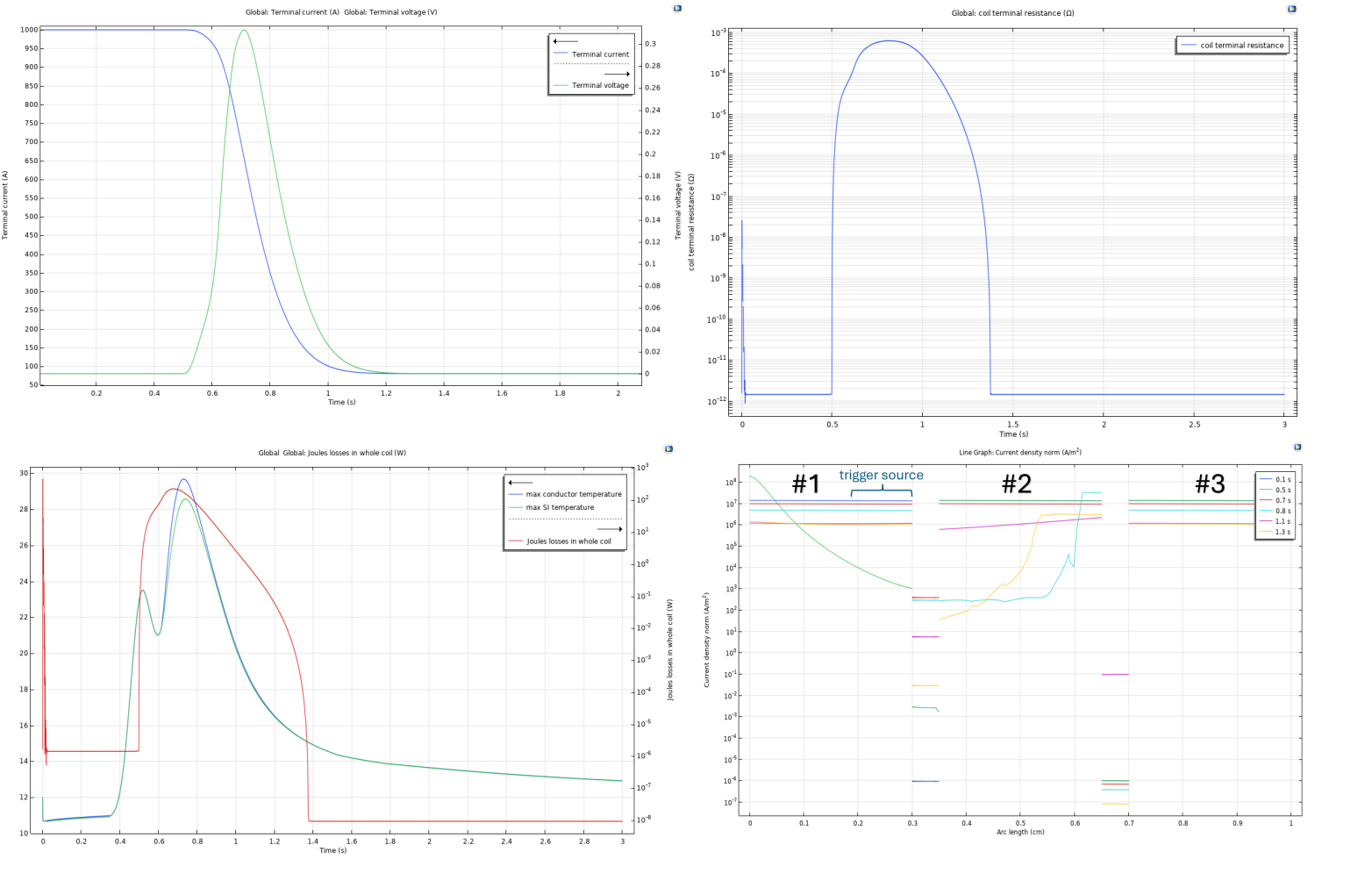}
  \caption{Reference quench simulation (3 turns, trigger on turn~1 at
    $t=0.5$~s): terminal current and voltage (top left); terminal
    resistance (top right); maximum conductor and SI temperature, and
    Ohmic losses (bottom left); current density along the winding at
    successive times, showing current sharing from turn~\#1 into the
    adjacent SI layer and later into turn~\#2 (bottom right).}
  \label{fig:model-quench-results}
\end{figure}

Figure~\ref{fig:model-quench-results} shows the result. After the
trigger, the conductor turns resistive and the coil current decays,
with the terminal resistance rising by about nine orders of magnitude.
The temperature shows two peaks: the trigger pulse, then a larger one
from Ohmic heating. The current-density profile along the winding
shows the mechanism of interest. Current leaves the hot conductor
segment and flows into the adjacent SI layer (turn~\#1). As heat
diffuses to turn~\#2, its SI layer also starts to conduct, and the
current-sharing path extends away from the hot spot.

\begin{figure}[tbp]
  \centering
  \includegraphics[width=0.7\textwidth]{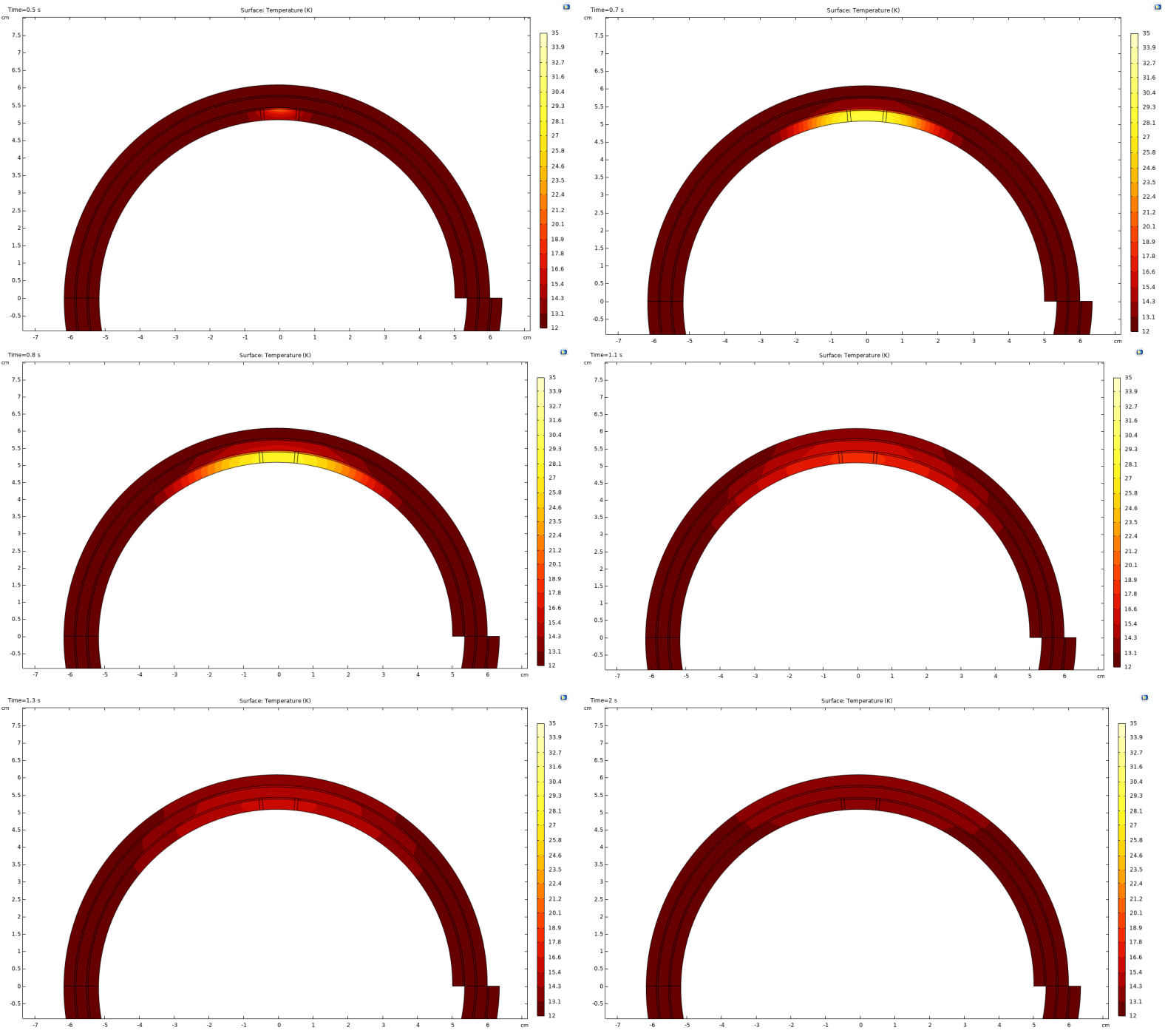}
  \caption{Temperature in the simulated 3-turn coil at
    $t = 0.5$, 0.7, 0.8, 1.1, 1.3 and 2.0~s: the hot spot at the trigger
    location on turn~\#1 broadens and migrates along the winding as
    current sharing through the SI layer proceeds, before the coil cools
    down once the 40~J stored energy is dissipated.}
  \label{fig:model-quench-tempprofile}
\end{figure}

With only 40~J stored, the coil cannot stay above $T_c$. The quench
self-extinguishes and the coil returns to within about a kelvin of its
initial temperature by $t = 2$~s. Realistic fusion-magnet quenches
involve orders of magnitude more energy. Runs at higher
stored energy, through larger inductance or current, did not yet
converge. Adaptive time-stepping, with fine steps only during the
transient, is the most likely fix. The reference case therefore shows
the current-sharing mechanism qualitatively, not its performance in a
realistic coil. Figure~\ref{fig:model-quench-tempprofile} shows the
simulated 3-turn geometry and the temperature map at successive times:
the hot spot at the trigger broadens along the winding as current
sharing proceeds, then the coil cools down.

\subsection{Outlook: refined coil model}
\label{sec:model-outlook}

A refined COMSOL model has since been developed and will be reported
separately \cite{ortino2026_iss}. It couples an axisymmetric
magnetostatic model (inductances, field) with an electro-thermal spiral
model, through an external circuit with dump resistor. The conductor
is REBCO with a field- and temperature-dependent critical current, and
the SI layer is a thin interface with a parametric
$\rho_{\mathrm{SI}}(T)$. First results on a 10-turn test coil at
$T_{op}=20$~K, with a local disturbance, already give quantitative
targets for the material. With a contrast of $10^5$ between insulating
and metallic resistivity and $T_{\mathrm{MIT}}=25$~K, below the
current-sharing temperature of the conductor ($\approx$37~K), the SI
coil protects itself like an NI coil. Its hot-spot temperature
($\approx$124~K) does not depend on the detection delay. For comparison,
the insulated coil with dump reaches 124~K with a 0.1~s delay and 214~K
with a 1~s delay. With $T_{\mathrm{MIT}}=50$~K, above the current-sharing
temperature, or with a contrast of only $10^4$, the SI layer opens only
locally and the hot spot rises to 146--201~K. The measured
V$_2$O$_3$--epoxy composite, with a contrast of only about 300 above
77~K, diverts practically no current. These results point to a contrast
of at least $10^5$ and a transition below the current-sharing
temperature, i.e.\ in the lower part of the 30--70~K window. They will
be confirmed on a full 200-turn pancake (about 1~MJ).

\section{Conclusions and outlook}
\label{sec:conclusions}

This exploratory study asked whether vanadium oxides can provide a
temperature-triggered smart-insulation layer for superconducting
coils. Such a layer would insulate in normal operation and conduct
only near a hot spot. For REBCO magnets, the transition should lie
between 30 and 70~K.

The main results are the following.
\begin{itemize}
  \item V$_2$O$_3$--epoxy specimens gave a robust reference. Their
    transition did not depend on the binder, was stable over thermal
    cycling and was unaffected by fields up to 15~T. From them we
    defined a practical acceptance criterion for new powders.
  \item TGA redox processing moved the effective transition of a
    multi-phase powder to $\approx$127~K, about 25--28~K below
    V$_2$O$_3$. Phase-pure products were not obtained. The V$_8$O$_{15}$
    fraction reached above 50\% in the latest batches.
  \item Multi-phase samples measured at ENEA showed a broad transition
    and a reproducible thermal hysteresis between warming and cooling,
    not explained by thermometry.
  \item Hydrothermal V$_3$O$_7\cdot$H$_2$O was far too resistive to serve
    as smart insulation. In the scarce literature, the $\sim$5~K anomaly
    of anhydrous V$_3$O$_7$ is antiferromagnetic ordering, not an MIT.
    This route is closed.
  \item The measured $c_p(T)$ of V$_2$O$_3$ differs by close to two
    orders of magnitude from the literature-based input of our first
    quench model.
  \item A first COMSOL model reproduces turn-to-turn current sharing
    through the SI layer, but only at low stored energy (40~J). A
    refined REBCO coil model \cite{ortino2026_iss} indicates that a
    resistivity contrast of at least $10^5$ and a transition below the
    current-sharing temperature make the SI coil protect itself like an
    NI coil, independently of the detection delay.
\end{itemize}

Several directions follow. The most promising short-term route is
doped V$_2$O$_3$. Titanium doping and oxygen excess are known to lower
its transition \cite{mcwhan1973,ueda1980}. This route keeps the
processing and in-coil experience already gained with V$_2$O$_3$. The
price is a smaller resistivity jump, which must be checked against the
acceptance criterion. In parallel, phase-pure V$_8$O$_{15}$ remains the
main stoichiometric target, at the upper edge of the HTS window. A
third route, reactive sputtering of V$_x$O$_y$ directly onto Cu, is
starting with the thin-film platform at IJL. It builds on established
VO$_2$ thin-film work. It would add strain as an independent tuning
lever, and it deposits the layer directly on a current-carrying
surface, which is closer to a real coil.

For LTS magnets, a smart-insulation layer could at most complement
voltage-based quench detection. Full self-protection appears excluded
because the normal zone spreads much faster than the layer can switch.
For HTS magnets, the slow quench dynamics match the thermal response of
an MIT layer. The full coil-scale electro-thermal study, with measured
material properties and realistic stored energies, will be reported in
a separate paper.

\ack{The authors thank H.~Monroe (MSc student, EPFL) for the COMSOL
simulations of the reference quench case, and F.~Oeberle (EPFL) for
technical support in specimen preparation and testing.}

\funding{This work was supported in part by the EUROfusion Consortium, via
the Euratom Research and Training Programme under Grant
101052200---EUROfusion, and in part by the Swiss State Secretariat for
Education, Research and Innovation (SERI). Views and opinions
expressed are however those of the author(s) only and do not necessarily
reflect those of the European Union or the European Commission. Neither
the European Union nor the European Commission can be held responsible
for them.}

\roles{\textbf{M.~Ortino:} Conceptualization, Methodology, Investigation,
Formal analysis, Data curation, Visualization, Funding acquisition,
Project administration, Writing -- original draft, Writing -- review \&
editing.
\textbf{D.~Uglietti:} Writing -- review \& editing.
\textbf{A.~Testino:} Writing -- review \& editing.
\textbf{N.~Casati:} Investigation (XRPD), Resources.
\textbf{A.~Steppke:} Investigation (PPMS measurements), Resources.
\textbf{W.~Wnetkowska:} Investigation (hydrothermal synthesis), Resources.
\textbf{A.~Masi:} Investigation (cryogenic $R(T)$ measurements at ENEA),
Resources.
\textbf{K.~Sedlak:} Project administration.}


\bibliographystyle{sust-smart-insulation}
\bibliography{references}

\providecommand{\newblock}{}
\begin{thebibliography}{10}
\expandafter\ifx\csname url\endcsname\relax
  \def\url#1{{\tt #1}}\fi
\expandafter\ifx\csname urlprefix\endcsname\relax\def\urlprefix{URL }\fi
\providecommand{\eprint}[2][]{\url{#2}}

\bibitem{wilson1983}
Wilson M~N 1983 {\em Superconducting Magnets\/} ({\em Monographs on
  Cryogenics\/} no~2) (Oxford: Clarendon Press / Oxford University Press) ISBN
  0198548052

\bibitem{hartwig2020}
Hartwig Z~S {\em et~al.\/} 2020 {\em Superconductor Science and Technology\/}
  {\bf 33} 11LT01

\bibitem{hartwig2024_tfmc}
Hartwig Z~S {\em et~al.\/} 2024 {\em IEEE Transactions on Applied
  Superconductivity\/} {\bf 34} 0600316

\bibitem{ortino2024_tasc}
Ortino M, Bykovskiy N, Sarasola X, Bruzzone P and Sedlak K 2024 {\em IEEE
  Transactions on Applied Superconductivity\/}

\bibitem{ortino2025_tasc_mit}
Ortino M and Bykovskiy N 2025 {\em IEEE Transactions on Applied
  Superconductivity\/}

\bibitem{mussa2024doped}
Mussa M~M, Noh H~S, Kwon D, Kim Y, Kim H~W, Jo Y~S, Kim S~W and Lee H 2024 {\em
  Chinese Journal of Physics\/} {\bf 87} 540--555

\bibitem{bhatia2018_review}
Bhatia R, Kumari K, Rani R, Suri A, Pahuja U and Singh D 2018 {\em Reviews in
  Physics\/} {\bf 3} 15--25

\bibitem{bhatia2018_weakT}
Bhatia R, Sameera I, Prasad V and Menon R 2018 {\em Materials Chemistry and
  Physics\/} {\bf 211} 258--263

\bibitem{guvvala2026_monap}
Guvvala N, Kim C~H, Cheetham P, Alam M~M, Manankandayalage C, Frey W, Prebys E,
  Tripathi M and Pamidi S~V 2026 {\em Advanced Manufacturing: Polymer \&
  Composites Science\/} {\bf 12}

\bibitem{galvin_varistor_2022}
Galvin T, Kirby G, Coll M {\em et~al.\/} 2022 Varistor insulation for {HTS}
  magnets

\bibitem{galvin2022_hotspot}
Galvin T, Kirby G, Coll M {\em et~al.\/} 2022 {\em IEEE Transactions on Applied
  Superconductivity\/}

\bibitem{galvin2023_sic4k}
Galvin T, Kirby G, Coll M {\em et~al.\/} 2023 {\em IEEE Transactions on Applied
  Superconductivity\/}

\bibitem{salazar2021_fiberoptic}
Salazar E~E, Badcock R~A, Bajko M, Castaldo B, Davies M, Estrada J, Fry V,
  Gonzales J~T, Michael P~C, Segal M, Vieira R~F and Hartwig Z~S 2021 {\em
  Superconductor Science and Technology\/} {\bf 34} 035027

\bibitem{brahlek2017}
Brahlek M {\em et~al.\/} 2017 {\em MRS Communications\/} {\bf 7} 27--52

\bibitem{hu2023}
Hu P {\em et~al.\/} 2023 {\em Chemical Reviews\/} {\bf 123} 4353--4415

\bibitem{kim2017_mit}
Kim H~W {\em et~al.\/} 2017 {\em IEEE Transactions on Applied
  Superconductivity\/}

\bibitem{jo2017_smartinsulation}
Jo Y~S {\em et~al.\/} 2017 {\em IEEE Transactions on Applied
  Superconductivity\/}

\bibitem{kim2019_racetrack}
Kim H~W {\em et~al.\/} 2019 {\em IEEE Transactions on Applied
  Superconductivity\/}

\bibitem{kim2022_racetrack300}
Kim H~W {\em et~al.\/} 2022 {\em IEEE Transactions on Applied
  Superconductivity\/}

\bibitem{yang2020_voxcoating}
Yang Z, Mart{\'i}nez A~C~A, Muley S~V, Wang X, Ji Q and Anders A 2020 {\em
  Journal of Applied Physics\/}

\bibitem{bonura2023contact}
Bonura M, Bovone G, Cayado P and Senatore C 2023 {\em IEEE Transactions on
  Applied Superconductivity\/} {\bf 33} 1--6

\bibitem{vaskuri2023}
Vaskuri A~K, Cur\'e B, Dudarev A and Mentink M 2023 {\em IEEE Transactions on
  Applied Superconductivity\/}

\bibitem{kawazoe2023_v8o15}
Kawazoe Y, Kanomata T and Note R 2023 V$_8$o$_{15}$ {\em High Pressure
  Materials Properties: Magnetic Properties of Oxides Under Pressure -- A
  Supplement to Landolt-B\"ornstein IV/22 Series\/} (Berlin, Heidelberg:
  Springer) pp 109--111

\bibitem{fisher2016_systematics}
Fisher B, Genossar J and Reisner G~M 2016 {\em Solid State Communications\/}

\bibitem{mit_review_2013_TODO}
Pergament A~L, Stefanovich G~B, Kuldin N~A and Velichko A~A 2013 {\em
  International Scholarly Research Notices\/} {\bf 2013} 960627

\bibitem{gardner1969}
Gardner R~A, Vlasse M and Wold A 1969 {\em Acta Crystallographica Section B:
  Structural Crystallography and Crystal Chemistry\/} {\bf 25} 781--787

\bibitem{matsuura1991}
Matsuura K, Wada T, Nakamizo T, Yamauchi H and Tanaka S 1991 {\em Physical
  Review B\/} {\bf 43} 13118

\bibitem{fazekas2007}
Fazekas P, Bari{\v s}i{\'c} N, K{\'e}zsm{\'a}rki I, Demk{\'o} L, Berger H,
  Forr{\'o} L and Mih{\'a}ly G 2007 {\em Physical Review B\/} {\bf 75} 035128

\bibitem{tanaka2007}
Tanaka A 2007 {\em Journal of Magnetism and Magnetic Materials\/} {\bf 310}
  898--900

\bibitem{li2009_v3o7}
Li C, Isobe M, Ueda H, Matsushita Y and Ueda Y 2009 {\em Journal of Solid State
  Chemistry\/} {\bf 182}

\bibitem{kuwamoto1981_v9o17}
Kuwamoto H, Otsuka N and Sato H 1981 {\em Journal of Solid State Chemistry\/}
  {\bf 36} 133--138

\bibitem{mcwhan1973}
McWhan D~B, Menth A, Remeika J~P, Brinkman W~F and Rice T~M 1973 {\em Physical
  Review B\/} {\bf 7} 1920

\bibitem{ueda1980}
Ueda Y, Kosuge K and Kachi S 1980 {\em Journal of Solid State Chemistry\/} {\bf
  31} 171--188

\bibitem{kengnizanguim2020_vox}
Kengni-Zanguim B, Joly L, Scheurer F, Ohresser P, Dayen J~F, Ulhaq-Bouillet C,
  Uzan J, Kundys B, Majjad H and Halley D 2020 {\em Applied Physics Letters\/}
  {\bf 116} 042404

\bibitem{willmott2013_addams}
Willmott P~R, Meister D, Leake S~J, Lange M, Bergamaschi A, B{\"o}ge M, Calvi
  M, Cancellieri C, Casati N, Cervellino A, Chen Q, David C, Flechsig U, Gozzo
  F, Henrich B, J{\"a}ggi-Spielmann S, Jakob B, Kalichava I, Karvinen P,
  Krempasky J, L{\"u}deke A, L{\"u}scher R, Maag S, Quitmann C, Reinle-Schmitt
  M~L, Schmidt T, Schmitt B, Streun A, Vartiainen I, Vitins M, Wang X and
  Wullschleger R 2013 {\em Journal of Synchrotron Radiation\/} {\bf 20}
  667--682

\bibitem{chine2011}
Chine M~K, Sediri F and Gharbi N 2011 {\em Materials Sciences and
  Applications\/} {\bf 2} 964--970

\bibitem{surikov2021}
Surikov V~I, Surikov V~I, Kuznetsova Y~V, Semenyuk N~A, Lyakh O~V and Prokudina
  N~A 2021 {\em Russian Physics Journal\/} {\bf 64} 376--380

\bibitem{andreev1978}
Andreev V~N, Chudnovskii F~A, Petrov A~V and Terukov E~I 1978 {\em physica
  status solidi (a)\/} {\bf 48} K153--K156

\bibitem{ortino2026_iss}
Ortino M {\em et~al.\/} 2026 Electro-thermal modelling of an {HTS} pancake coil
  with temperature-triggered smart insulation in preparation, to be presented
  at ISS2026

\end{thebibliography}

\end{document}